\documentclass[acmlarge]{acmart}
\AtBeginDocument{%
  }

\setcopyright{acmlicensed}
\acmDOI{XXXXXXX.XXXXXXX}

\acmJournal{HEALTH}
\usepackage{algorithm}
\usepackage{array}
\usepackage{textcomp}
\usepackage{stfloats}
\usepackage{url}
\usepackage{verbatim}
\usepackage{graphicx}
\usepackage{tabularx}
\usepackage{subcaption}
\usepackage{tikz}
\usepackage{xcolor}
\usepackage{mathrsfs}%
\usepackage{comment}
\usepackage{hyperref}
\usepackage{algpseudocode}
\usepackage{wrapfig}
\usepackage{multirow}
\usepackage{orcidlink}
\usepackage{soul}
\usepackage{bm}

\usepackage{tikz}
\usepackage{pgfplots}
\usepackage{filecontents}
\usepackage{booktabs}

\DeclareUnicodeCharacter{2019}{\textquotesingle} % ’
\DeclareUnicodeCharacter{00B4}{\textquotesingle} % ´
\DeclareUnicodeCharacter{2032}{\ensuremath{^{\prime}}} % ′

\begin{document}

\title{Data-Efficient Psychiatric Disorder Detection via Self-supervised Learning on Frequency-enhanced Brain Networks}


\author{Mujie Liu}
\email{mujie.liu@ieee.org}
\orcid{0009-0002-0879-7168}
\affiliation{%
\department{Institute of Innovation, Science and Sustainability}
  \institution{Federation University Australia}
  \city{Ballarat}
  \state{VIC 3353}
  \country{Australia}
}

\author{Mengchu Zhu}
\email{22020090107@pop.zjgsu.edu.cn}
\orcid{0009-0001-3438-4394}
\author{Qichao Dong}
\email{23020090088@pop.zjgsu.edu.cn}
\orcid{0009-0008-5986-0124}
\affiliation{%
\department{School of Information and Electronic Engineering}
  \institution{Zhejiang Gongshang University}
  \city{Hangzhou}
  \state{Zhejiang 315104}
  \country{China}
}

\author{Ting Dang}
\email{ting.dang@unimelb.edu.au}
\orcid{0000-0003-3806-1493}
\authornote{Corresponding author.}
\affiliation{%
\department{School of Computing and Information Systems}
  \institution{The University of Melbourne}
  \city{Melbourne}
  \state{VIC 3052}
  \country{Australia}
}

\author{Jiangang Ma}
\email{j.ma@federation.edu.au}
\orcid{0000-0002-8449-7610}
\affiliation{%
\department{Institute of Innovation, Science and Sustainability}
  \institution{Federation University Australia}
  \city{Ballarat}
  \state{VIC 3353}
  \country{Australia}
}

\author{Jing Ren}
\email{S4065838@student.rmit.edu.au}
\orcid{0000-0003-0169-1491}
\author{Feng Xia}
\email{f.xia@ieee.org}
\orcid{0000-0002-8324-1859}
\affiliation{%
\department{School of Computing Technologies}
  \institution{RMIT University}
  \city{Melbourne}
  \state{VIC 3000}
  \country{Australia}
}

\renewcommand{\shortauthors}{M. Liu et al.}

\begin{abstract}
Psychiatric disorders involve complex neural activity changes, with functional magnetic resonance imaging (fMRI) data serving as key diagnostic evidence. However, data scarcity and the diverse nature of fMRI information pose significant challenges. While graph-based self-supervised learning (SSL) methods have shown promise in brain network analysis, they primarily focus on time-domain representations, often overlooking the rich information embedded in the frequency domain. 
To overcome these limitations, we propose \textbf{F}requency-\textbf{E}nhanced \textbf{Net}work (FENet), a novel SSL framework specially designed for fMRI data that integrates time-domain and frequency-domain information to improve psychiatric disorder detection in small-sample datasets. FENet constructs multi-view brain networks based on the inherent properties of fMRI data, explicitly incorporating frequency information into the learning process of representation. Additionally, it employs domain-specific encoders to capture temporal-spectral characteristics, including an efficient frequency-domain encoder that highlights disease-relevant frequency features. Finally, FENet introduces a domain consistency-guided learning objective, which balances the utilization of diverse information and generates frequency-enhanced brain graph representations. Experiments on two real-world medical datasets demonstrate that FENet outperforms state-of-the-art methods while maintaining strong performance in minimal data conditions. Furthermore, we analyze the correlation between various frequency-domain features and psychiatric disorders, emphasizing the critical role of high-frequency information in disorder detection.  
\end{abstract}

\begin{CCSXML}
<ccs2012>
   <concept>
       <concept_id>10010405.10010444.10010447</concept_id>
       <concept_desc>Applied computing~Health care information systems</concept_desc>
       <concept_significance>500</concept_significance>
       </concept>
   <concept>
       <concept_id>10010405.10010444.10010087.10010091</concept_id>
       <concept_desc>Applied computing~Biological networks</concept_desc>
       <concept_significance>500</concept_significance>
       </concept>
   <concept>
        <concept_id>10010147.10010257.10010258</concept_id>
        <concept_desc>Computing methodologies~Learning paradigms</concept_desc>
        <concept_significance>300</concept_significance>
        </concept>
   <concept>
       <concept_id>10010147.10010257.10010293.10010319</concept_id>
       <concept_desc>Computing methodologies~Learning latent representations</concept_desc>
       <concept_significance>300</concept_significance>
       </concept>
 </ccs2012>
\end{CCSXML}

\ccsdesc[500]{Applied computing~Health care information systems}
\ccsdesc[500]{Applied computing~Biological networks}
\ccsdesc[300]{Computing methodologies~Learning paradigms}
\ccsdesc[300]{Computing methodologies~Learning latent representations}

\keywords{Psychiatric disorder detection, data scarcity, brain networks, self-supervised learning, frequency-enhanced brain graph representation}


\maketitle

\section{Introduction}

Psychiatric disorders, including depression, autism spectrum disorder (ASD), and attention deficit hyperactivity disorder (ADHD), are complex mental health conditions characterized by distinct neurological and behavioral patterns on brain networks \cite{pathak2024does, muetunda2025ai, du2025improving, bayat2025reduced}. Functional magnetic resonance imaging (fMRI) measures brain activity and connectivity by detecting blood-oxygen-level-dependent (BOLD) signal fluctuations \cite{arthurs2002well, du2024survey}, offering valuable insights into neural function and dysfunction. Consequently, fMRI has been widely used to investigate functional abnormalities of brain and neurodevelopmental disorders, facilitating diagnosis and discovery of biomarkers \cite{menon2019comparison, xia2017functional}.

\begin{figure}[t]
        \centering
        \begin{subfigure}[b]{0.45\textwidth} % 调整宽度
            \centering
            \includegraphics[width=\textwidth]{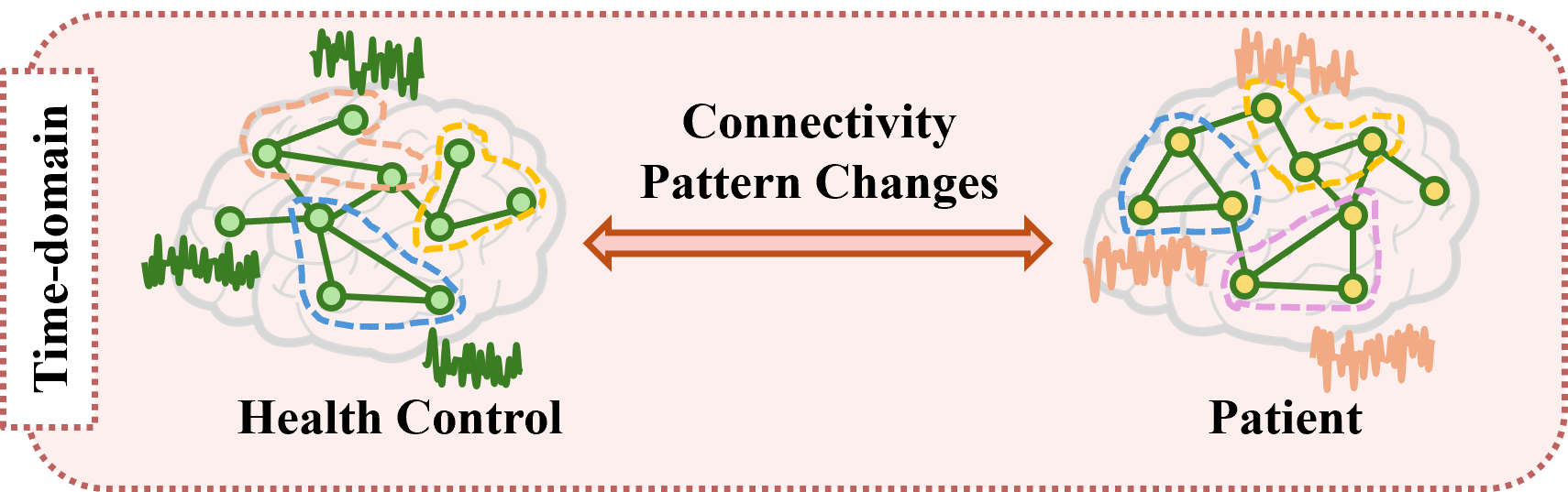}
            \caption{Time-domain brain networks for the health control and patient}
            \label{time-domain}
        \end{subfigure}
        \hspace{5pt}
        \begin{subfigure}[b]{0.45\textwidth} % 调整宽度
            \centering
            \includegraphics[width=\textwidth]{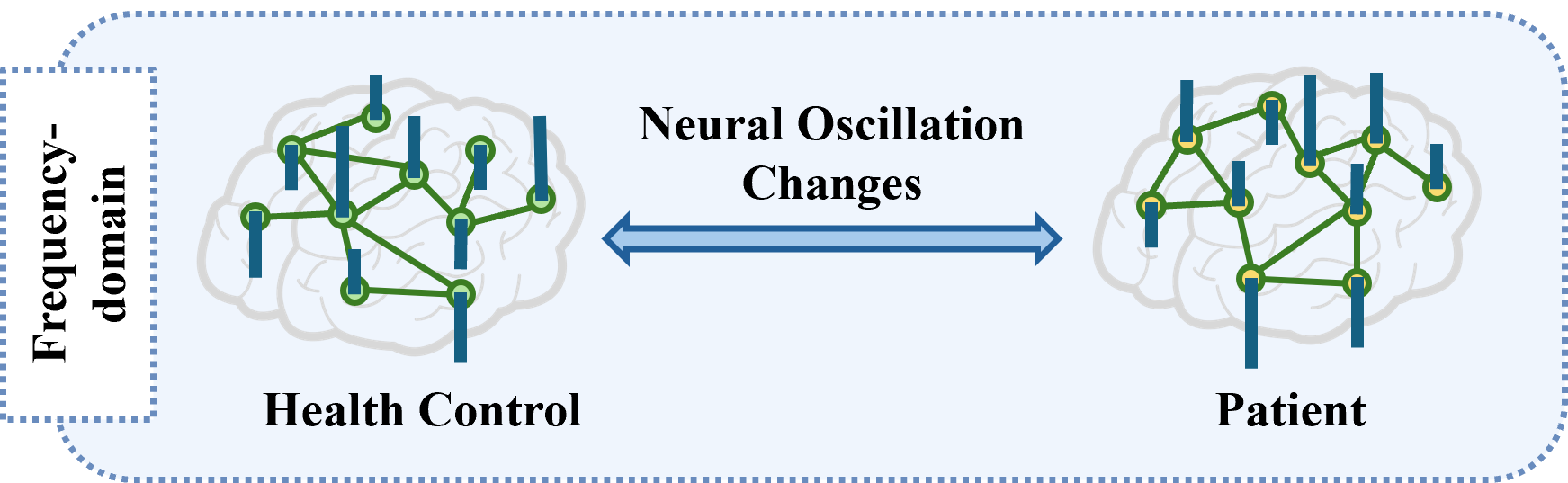}
            \Description{Frequency-domain brain networks for the health control and patient}
            \caption{Frequency-domain brain networks for the health control and patient}
            \label{frequency-domain}
        \end{subfigure}

        \caption{Illustration of fMRI information diversity in psychiatric disorder detection. Brain network analysis of health control and patient fMRI data reveals that disease states correlate with changes in connection patterns in the time domain and alterations in neural oscillations in the frequency domain.}
        \end{figure}

However, effectively analyzing fMRI data for psychiatric disorder detection remains challenging due to data scarcity and the complexity of its diverse information \cite{sartorius2017comorbidity, hutchison2013dynamic, dvornek2017identifying, santosrole}. 
A major hurdle is the difficulty in collecting large, high-quality datasets, as psychiatric diagnosis often requires long-term clinical observation and expert evaluation \cite{regier2013dsm, hyman2020executive}. Additionally, high neuroimaging costs and ethical constraints further limit data availability, resulting in small, imbalanced datasets with inconsistent preprocessing, making it difficult to train robust diagnostic models. Beyond data limitations, fMRI inherently contains both time-domain (temporal) and frequency-domain (spectral) information. However, integrating these complementary domains is challenging due to their distinct representations and the complexity of brain signals. Specifically, time-domain features capture BOLD signal fluctuations, modeling the functional connectivity of brain networks. As shown in Figure~\ref{time-domain}, changes in connectivity patterns reflect transitions from healthy to diseased states~\cite{ thottupattu2024method}. 
In contrast, frequency-domain features capture neural oscillations, revealing intrinsic brain rhythms and disease-specific frequency changes~\cite{huang2016graph, sasai2021frequency}. For example, in ASD, low-frequency correlations decline with age, while high-frequency signals highlight disorder traits~\cite{jafadideh2022rest, karavallil2023amplitude}. Figure~\ref{frequency-domain} shows differences in spectral coefficients between healthy and patient groups. 
This diversity complicates the diagnostic process, as existing models often struggle to capture the full spectrum of information. Thus, time-frequency joint modeling is crucial for accurate psychiatric disorder detection as integrating frequency information provides complementary insights and improves data efficiency, particularly in small-sample settings.

Recent advancements have increasingly focused on graph learning-based approaches, with graph neural networks (GNNs) \cite{9416834, peng2024learning, luo2024graph} emerging as a powerful tool for fMRI data analysis. Given their ability to model non-Euclidean structures, GNNs are particularly well-suited for brain network construction and extracting complex representations that reflect brain function \cite{ liu2024motif, bi2023community, luo2024knowledge}. For example, BrainGNN \cite{li2021braingnn} utilizes ROI-aware GNN to facilitate biomarker discovery, while BrainGB \cite{cui2022braingb} establishes a benchmark by summarizing the brain network construction pipelines and modularizing the GNN process for brain network analysis. However, these models heavily rely on large labeled datasets, which are often scarce in psychiatric disorder research. To mitigate data scarcity, recent studies have explored self-supervised learning (SSL) on graphs \cite{liu2022graph, kumar2022contrastive} to enable representation learning from unlabeled data. Some SSL methods select positive and negative sample pairs in augmented brain networks, which encourages better representation learning at the node level \cite{zhang2023gcl, luo2024interpretable, xu2024contrastive, zong2024new}. Other approaches leverage the temporal characteristics of fMRI data by constructing augmented views across different time windows, aiming to capture the inherent temporal dependencies in the representations~\cite{peng2022gate, wang2023unsupervised}.

Despite their promising performance, existing SSL methods encounter several limitations when applied to fMRI data, primarily due to their neglect of frequency information, leading to suboptimal data efficiency. 
First, most SSL approaches rely heavily on time-domain data augmentation strategies, such as edge or attribute perturbation~\cite{zhang2023gcl, luo2024interpretable, xu2024contrastive, zong2024new, yang2023ts}. However, these operations often disrupt the structural integrity of brain networks~\cite{peng2022gate}, potentially distorting their inherent time-frequency properties and leading to inauthentic representations. To address this, we construct biologically consistent multi-view representations of fMRI data across both time and frequency domains. 
Second, existing SSL methods typically adopt time-domain encoders, which face inherent limitations in modeling the complex brain functions and intrinsic cyclic patterns that are better captured in the frequency domain~\cite{zhang2022self, liu2024frequency}. As a result, these methods often fail to exploit frequency-related neural activity, overlooking critical features that are particularly valuable in limited-sample scenarios. To remedy this, we propose carefully designed pretext tasks to incorporate frequency-domain analysis and facilitate effective time-frequency joint modeling. 
Third, most traditional SSL objectives are designed to learn representations within a single time domain, lacking the capability to align cross-domain features. This makes it difficult to effectively integrate complementary information from both time and frequency domains, resulting in incomplete utilization of the available data. To address this, we refine the SSL objective to support cross-domain representation alignment, enabling the model to learn joint time-frequency embeddings that are more comprehensive and discriminative. 

In this paper, we propose FENet, a novel SSL framework that integrates time- and frequency-domain analysis to capture diverse information from fMRI data, enhancing data efficiency and improving diagnostic performance in limited-sample scenarios.
Specifically, FENet focuses on biologically consistent modeling of multi-view brain networks in time and frequency domains, offering complementary perspectives on fMRI data. Meanwhile, FENet employs dedicated domain-specific encoders: a time-domain GNN (TGNN) and a frequency-domain GNN (FGNN), which collaboratively perform time-frequency joint modeling, fully leveraging the complementary information to enhance data efficiency. Notably, the FGNN incorporates adaptive graph filtering to selectively capture informative frequency components, while the integration of Fourier graph operator (FGO) layers~\cite{huang2016graph} reduces computational complexity to a log-linear scale, enabling efficient and scalable frequency-domain analysis in complex brain networks.
Furthermore, FENet employs a learning objective based on canonical correlation analysis (CCA)~\cite{zhang2021canonical}, guided by the domain consistency concept to capture domain-invariant features and optimize both time- and frequency-domain representations. This enables FENet to generate frequency-enhanced brain network representations for effective psychiatric disorder detection.

	\begin{itemize}
		\item We propose FENet, a novel non-contrastive SSL framework for data-efficient psychiatric disorder detection, which is the first to integrate time- and frequency-domain analysis for learning brain network representations. 
        \item 
        We introduce a novel data augmentation strategy that constructs brain graphs in both the time and frequency domains, ensuring biological consistency with real brain networks while effectively modeling the diverse information embedded in fMRI data. 

        \item 
        We design a frequency-domain information encoder that enables the selective learning of frequency components for psychiatric disorders using the proposed graph filter and FGO layers.

		\item  We conduct extensive experiments on two real-world medical datasets, demonstrating that FENet achieves state-of-the-art performance in limited-sample settings. Additionally, frequency domain analysis reveals a significant correlation between high-frequency components and psychiatric disorders.

	\end{itemize}

The structure of the paper is as follows: Section~\ref{Related Works} reviews related work. Then, the preliminary definitions and notations are introduced in Section~\ref{preliminaries}. Section~\ref{The Proposed Method} presents the details of the proposed FENet method. Section~\ref{Experiments} outlines the experimental setup, followed by the results and their discussion in Section~\ref{Results and Discussion}. Lastly, Section~\ref{Conclusion} concludes the paper.

\section{Related Work}
\label{Related Works}
In this section, we introduce the most relevant works from three perspectives: graph representation learning for brain signals, SSL on graphs, and frequency-based time series analysis. 
    
 	\subsection{Graph Representation Learning for fMRI Data}
    \label{2.1}
	Graph representations of brain signals are widely recognized for their effectiveness in capturing the underlying dynamics of brain functions using GNNs~\cite{10388338}. In these models, nodes typically correspond to regions of interest (ROIs), while edges that quantify functional connectivity between different ROIs are generally defined by the correlations between their respective brain signals. Building on this concept, BrainGNN~\cite{li2021braingnn} introduces a novel ROI-aware GNN with an ROI-selection pooling layer highlighting the salient brain ROIs. 
    Recently, BrainGB~\cite{cui2022braingb} further established a benchmark that formulates the implementation of GNN models for brain network analysis, systematically demonstrating how different GNN mechanisms influence brain network modeling and performance. 
    Despite the promising results achieved by these approaches in capturing the features of brain networks, they are still limited by the scarcity of labeled data. To address this challenge, SSL-based methods have emerged as an effective solution, which involves pre-training with limited data to learn effective representations, followed by fine-tuning to improve generalization and accuracy. For instance, A-GCL~\cite{zhang2023gcl} and SF-GCL~\cite{peng2024stage} investigate contrastive learning for brain signals by introducing random edge manipulation to generate augmented data views. Besides, BraGCL~\cite{luo2024interpretable} improves model interpretability by perturbing unimportant nodes and edges. 
    Further, GATE~\cite{peng2022gate} enhances the pair constructions for SSL. 
    UCGL~\cite{wang2023unsupervised} extends this approach by integrating a spatio-temporal graph convolution technique. While these SSL-based methods have improved the representation learning of brain signals, they focus exclusively on time-domain features. Consequently, they fail to comprehensively capture information in the frequency domain and their correlation with psychiatric disorders. To bridge this gap, our approach explores the frequency-domain characteristics of brain signals and integrates them with time-domain details, enabling a more comprehensive representation. 
	
	\subsection{Self-supervised Learning on Graphs}

	\label{2.2}
    Existing SSL-based methods \cite{ren2023GL, 9709096, tang2022contrastive} typically construct pairs from the input through augmentation and learn the representations by maximizing agreement between positive (similar) pairs while minimizing the similarity between negative (dissimilar) pairs. 
    The effectiveness of SSL heavily depends on negative pair setting and the learning objective functions. Data augmentation techniques are widely used in the construction of negative samples. For example, GraphCL \cite{you2020graph} employs edge perturbation, node feature masking, or subgraph sampling, while GCA \cite{zhu2021graph} integrates structural and attribute information for adaptive augmentation. Additionally, some methods leverage graph spectral features, such as SpCo \cite{liu2022revisiting} and SFA \cite{zhang2023spectral}, to guide the edge perturbation process, which emphasizes high-frequency information related to graph structures. However, these random perturbations can inevitably disrupt intrinsic biological structures in brain data, leading to implausible outcomes. In contrast, approaches like GATE \cite{peng2022gate} and UCGL \cite{wang2023unsupervised} create augmented views based on the time-series properties, focusing on learning temporal dependencies instead of applying random perturbations. However, they still overlook a detailed examination of frequency information.
    To address this issue, we construct graphs that preserve the biological features while also incorporating the temporal-spectral characteristics of fMRI data for better multiple graph view construction, both in the time and frequency domains. 
    
    In terms of the learning objective functions, they are generally categorized into contrast-based and similarity-based approaches. Contrast-based methods, such as GraphCL \cite{you2020graph} and GCA \cite{zhu2021graph}, require careful design of positive and negative sample pairs. 
    However, constructing negative pairs is generally challenging, especially in data-limited conditions \cite{wu2024statiocl}, which impacts their effectiveness. 
    On the other hand, the similarity-based methods do not rely on selecting negative pairs. 
    Examples include CCA-SSG \cite{zhang2021canonical}, which simplifies SSL via CCA that aims at maximizing the correlation between two augmented views \cite{andrew2013deep}. Similarly, GATE \cite{peng2022gate} utilizes CCA for brain signal analysis to capture the correlation between representations of adjacent augmented views within a specified time window. However, no prior work has explored using CCA to align time- and frequency-domain representations. We are the first to introduce the concept of domain consistency to guide CCA in maximizing the correlation of representation from different domains.
    
	\subsection{Frequency-Based Time Series Analysis}
	
	\label{2.3}

    Brain signals are inherently time series data, with frequency information capturing oscillatory patterns and functional connectivity that play critical roles in many diseases \cite{huang2016graph,sasai2021frequency}. Consequently, incorporating frequency information is essential in representation learning. Existing methods have made significant progress in this area. Techniques such as the Fourier Transform decompose time series into frequency components, revealing patterns and periodicities at various scales. These methods have been widely applied in time series analysis, including FEDformer \cite{zhou2022fedformer}, Fredformer \cite{piao2024fredformer}, and FreTS \cite{yi2024frequency}. A few studies have incorporated frequency information in graph learning, such as StemGNN~\cite{cao2020spectral} and FourierGNN~\cite{yi2023fouriergnn}. 
    However, these approaches focus solely on frequency-domain features while neglecting the complementary information present in the time domain. This limitation restricts their applicability, particularly in scenarios involving limited labeled fMRI datasets, where fully leveraging both domains is crucial for effective representation learning.
    
    BTSF \cite{yang2022unsupervised} is the first to integrate time-frequency affinity into the SSL paradigm, enhancing the discriminative power of frequency feature fusion. Furthermore, Zhang et al. \cite{zhang2022self} introduce a pre-training model to uncover general properties connecting time and frequency information and leverage SSL to capture these properties. 
    Despite these advancements, there is a lack of studies examining the time-frequency integration in brain network modeling. Integrating frequency information into the traditional SSL approaches for brain networks remains challenging due to a lack of prior knowledge on biologically relevant frequency features and the high computational cost of extracting frequency-domain features from graphs. To overcome these issues, FENet introduces carefully designed pretext tasks that examine the contribution of different frequency components to disease detection, facilitating in-depth frequency analysis and enhancing representation learning for brain networks. Additionally, it incorporates efficient FGO layers, reducing the computational complexity to log-linear complexity.

    \section{Preliminaries}\label{preliminaries}
    In this section, we provide preliminaries of this paper, including spectral graph theory, the concept of domain consistency, and the criterion of CCA. Table~\ref{tab: notations} summarizes the notations that are frequently used in this paper, where bold lowercase letters (e.g., $\mathbf{x}$), bold uppercase letters (e.g., $\mathbf{X}$), and calligraphic fonts (e.g., $\mathcal{V}$) are used to denote vectors, matrices, and sets, respectively.

\begin{table}[t]
    \centering
    \caption{Notations and explanations related to the FENet framework. The four blocks (from top to bottom) display the notation for variables related to brain graphs, the graph Fourier transformer, SSL, and hyperparameters of FENet, respectively. }
    \label{tab: notations}
    \begin{tabularx}{\columnwidth}{c X} % Adjusts column width automatically

        \toprule
        \textbf{Notation} & \textbf{Description} \\
        \midrule
        $\mathcal{G}_\text{T}= (\mathcal{V}, \mathcal{E}, \mathbf{X}_\text{T}, \mathbf{A})$ & Time-domain brain graph \\
        $\mathcal{G}_\text{F}= (\mathcal{V}, \mathcal{E}, \mathbf{X}_\text{F}, \mathbf{A})$ & Frequency-domain brain graph \\
        $\mathcal{V}$ & The node set of the brain graph\\
        $\mathcal{E}$ & The edge set of the brain graph\\
        $\mathbf{A} \in \mathbb{R}^{N \times N}$ & Adjacency matrix of the brain graph\\
        $\mathbf{X}_\text{T}\in \mathbb{R}^{N \times D} $ & Time-domain node feature matrix with $D$-dimensional BOLD signals, $\mathbf{X}_\text{T}=[\mathbf{x}_\text{T}^1, \dots, \mathbf{x}_\text{T}^N], \mathbf{x}_\text{T}\in \mathbb{R}^{D}$\\
        $\mathbf{X}_\text{F}\in \mathbb{R}^{N \times D} $ & Frequency-domain node feature matrix with $D$-dimensional spectral coefficient, $\mathbf{X}_\text{F}=[\mathbf{x}_\text{F}^1, \dots, \mathbf{x}_\text{F}^N], \mathbf{x}_\text{F}\in \mathbb{R}^{D}$\\
        $N$ & Number of nodes (ROIs) in the brain graph \\
        
        \midrule
        $\mathbf{D} \in \mathbb{R}^{N \times N}$ & Diagonal matrix of the graph \\
        $\mathbf{L} \in \mathbb{R}^{N \times N}$ & Laplacian matrix of the graph \\ 
        $\mathbf{I} \in \mathbb{R}^{N \times N}$ & Identity matrix \\
        $\mathbf{U} \in \mathbb{R}^{N \times N}$ & Orthonormal matrix of eigenvectors \\
        $\mathbf{\Lambda} \in \mathbb{R}^{N \times N} $ & Diagonal matrix with eigenvalues, $\mathbf{\Lambda}=\text{diag}(\mathbf{\lambda})$\\

        \midrule
        $\mathbf{S}\in \mathbb{R}^{ K \times K}$  &Fourier graph operator \\
        $\mathbf{H} \in \mathbb{R}^{N \times N}$ & Graph filter matrix \\
        $\mathbf{W} \in \mathbb{R}^{N \times N}$ & Learnable parameter matrix \\
        $\mathbf{Z}_\text{T},\mathbf{Z}_\text{F}, \mathbf{Z}_\text{TF} \in \mathbb{R} ^{N \times D }$ & Time-domain, frequency-domain and fused brain graph representations\\

        \midrule
        $K_{\text{L}}, K_{\text{H}} $ & Parameters for frequency threshold setting \\
        $\gamma,\beta$ & Trade-off coefficient for learning objective\\
       
        \bottomrule
    \end{tabularx}
\end{table}
    
       \subsection{The Spectral Graph Theory} 
   In time series data, frequency components are characterized by periodic patterns in the signal, capturing changes or rates of oscillation over time. Similarly, in graph-structured data, the frequency component is defined by variations in the signal among connected nodes, reflecting how the signal changes between the neighboring nodes. Low-frequency information in graph-structured data suggests that signals among neighboring nodes are similar or aligned, while high-frequency information indicates greater variability or differences in the signal between connected nodes. Graph Fourier Transform (GFT) is commonly adopted to represent frequency information in graph data~\cite{shuman2013emerging}.

    For a given graph $\mathcal{G} = (\mathcal{V}, \mathcal{E}, \mathbf{X}, \mathbf{A})$, where $\mathcal{V}$ and $\mathcal{E}$ represent the set of nodes and edges, $\mathbf{X}\in \mathbb{R}^{N \times D}$ denotes the graph signals, $\mathbf{A}\in \mathbb{R}^{N \times N}$ is the adjacency matrix encoding the functional connectivity between node pairs, and $N$ is the number of nodes, the GFT aims to map the graph signal into the spectral domain. This is achieved through eigenvector-based feature decomposition of the graph Laplacian matrix, defined as:
    \begin{equation}
    \mathbf{L} = \mathbf{D} - \mathbf{A}
    \end{equation}
    where \(\mathbf{D}\in \mathbb{R}^{N \times N}\) is a diagonal matrix encoding the degree of each node. The Laplacian matrix \(\mathbf{L}\in \mathbb{R}^{N \times N}\) is real, symmetric, and positive semi-definite, allowing for eigendecomposition:
    \begin{equation}
    \mathbf{L} = \mathbf{U} \mathbf{\Lambda} \mathbf{U}^\top,
    \end{equation}
    where \(\mathbf{U}\in \mathbb{R}^{N \times N}\) is the orthonormal matrix of eigenvectors and \(\mathbf{\Lambda}=\text{diag}(\mathbf{\lambda})\) is the diagonal matrix of eigenvalues. The eigenvalues of the graph Laplacian, indexed as
    $0 < \lambda_1 \le \dots \le \lambda_{N}$, are referred to as the graph frequency components. Their associated eigenvectors constitute the Fourier modes. These components provide a foundation for analyzing the graph signals in the spectral domain. 
    Thus, for the given graph signal \(\mathbf{X} \), using the GFT, one can transform \(\mathbf{X} \) from the spatial domain to the spectral (frequency) domain as follows:
    \begin{equation}
    {\mathbf{X}_{\text{F}}} = \mathbf{U}^\top \mathbf{X}.
    \end{equation}
    As $\mathbf{U}$ is the orthonormal matrix, the inverse graph Fourier transform (IGFT) is given by:
    \begin{equation}
    \mathbf{X} = \mathbf{U} \,{\mathbf{X}_{\text{F}}}.
    \end{equation}
    \par

    To preserve certain information of graph data in the frequency domain, the process involves filtering, which is achieved by first transforming the signal using the GFT to the frequency domain, conducting the filtering, and then converting it back to the time domain as follows: 
    \begin{equation}
    \widetilde{\mathbf{X}} = \mathbf{U}\, \mathbf{H}\, \mathbf{U}^\top \mathbf{X}.
    \end{equation}
    Here, the \(\mathbf{H}\in \mathbb{R}^{N \times N}\) is a diagonal filtering matrix that offers flexibility in analyzing specific frequency components, and $\widetilde{\mathbf{X}}$ is the filtered feature matrix. For example, larger eigenvalues are associated with higher variance eigenvectors, which can be interpreted as higher graph frequencies~\cite{jafadideh2022rest}. Consequently, the frequency information of graph signals can be selectively retained by applying filters $\mathbf{H}$.
    
	\subsection{Domain Consistency Concept}

    The concept of domain consistency between time and frequency information is first introduced by Zhang et al.~\cite{zhang2022self} in the context of SSL for time series modeling. It aims to align the time- and frequency-domain representations by emphasizing the domain invariant feature learning in latent space. 
    For a given time series data $\mathbf{X} = [\mathbf{x}^1, \ldots, \mathbf{x}^ N]$, it is assumed that the latent representations of the time domain $\mathbf{z}^ i_\text{T}$ and frequency domain $\mathbf{z}^ i_\text{F}$ derived from the same original time series $\mathbf{x}^i, i \in [1, N]$ should closely align, while representations from different original time series $\mathbf{x}^j, j \in [1, N]$ should be pushed farther apart. This concept guides the learning objective of effectively learning domain-invariant features, thereby achieving the integration of information from both domains.  
    
	\subsection{Canonical Correlation Analysis}
	CCA~\cite{andrew2013deep} is a statistical technique used to analyze the correlations between two sets of variables. It aims to find a linear transformation that maximizes the correlation between the transformed variables.  It has been widely applied in multi-view fusion and fMRI analysis~\cite{zhang2021canonical,zhuang2020technical,peng2022gate}, to enhance the understanding of relationships between different views of a graph. Specifically, assuming $\mathcal{G}_a$ and $\mathcal{G}_b$ are two views of the graph, it optimizes the correlation: 
	\begin{equation}
		\begin{aligned}
			\underset{\theta_1,\theta_2}{\text{max}}\mathcal{L}_{CCA} := \mathbb{E}_{\mathcal{G}_a,\mathcal{G}_b}\left[f_{\theta_1}^\top(\mathcal{G}_a) f_{\theta_2}(\mathcal{G}_b)\right], \\
			\text{s.t.} \operatorname{Cov}\left(f_i(\mathcal{G}_j),f_i(\mathcal{G}_j)\right) = \mathbf{I},(i=\theta_1,\theta_2;j=a,b),
		\end{aligned}\label{CCA}
	\end{equation}
	where $f_{\theta_1}(\cdot)$ and $f_{\theta_2}(\cdot)$ are two feedforward neural networks normally sharing parameters, $\operatorname{Cov}(\cdot , \cdot)$ represents the covariance matrix and $\mathbf{I}$ is an identity matrix. The learned feedforward neural networks provide mappings that align the representations of nodes in the two views. Since CCA does not rely on the non-trivial construction of negative sample pairs, it offers greater flexibility for graph-level representation learning in SSL settings while also being scalable to large and complex graphs. 
    However, traditional CCA is computationally expensive. To address this issue, a recent study proposed soft CCA~\cite{chang2018scalable}, which relaxes the hard decorrelation constraints by applying the Lagrangian theorem within convex optimization.
    The formulation of soft CCA is defined as follows:
	\begin{equation}\label{CCA_L}
		\begin{aligned}
			\underset{\theta_1,\theta_2}{\text{min}}\mathcal{L}_{DIST}\left(f_{\theta_1}(\mathcal{G}_a),f_{\theta_2}(\mathcal{G}_b)\right) + \\
			\beta \left(\mathcal{L}_{SDL}\left(f_{\theta_1}(\mathcal{G}_a)\right) + \mathcal{L}_{SDL}\left(f_{\theta_2}(\mathcal{G}_b)\right)\right),
		\end{aligned}
	\end{equation}
	where $\mathcal{L}_{DIST}(\cdot,\cdot)$ is used to measure the correlation between two view representations, $ \mathcal{L}_{SDL}(\cdot)$, referred to as random decorrelation loss, computes the $L_1$ distance between each representation and the identity matrix functioning as the regularization, and the $\beta$ is a trade-off coefficient. The decorrelation term can be viewed as a soft constraint, as the loss is minimized rather than strictly enforced to zero. By jointly optimizing the decorrelation loss $ \mathcal{L}_{SDL}(\cdot)$ alongside other objectives, such as minimizing the distance between views in the embedding space, the model can explore a broader set of globally optimal solutions. Therefore, soft CCA can effectively enhance the correlation 
    between two graph views after transformation, while also enforcing regularization to reduce computational costs and avoid overfitting in SSL settings.
    
    \section{The Proposed Method}
    \label{The Proposed Method}
\begin{figure*}[t!]
        \centering
        \includegraphics[width=0.9\textwidth]{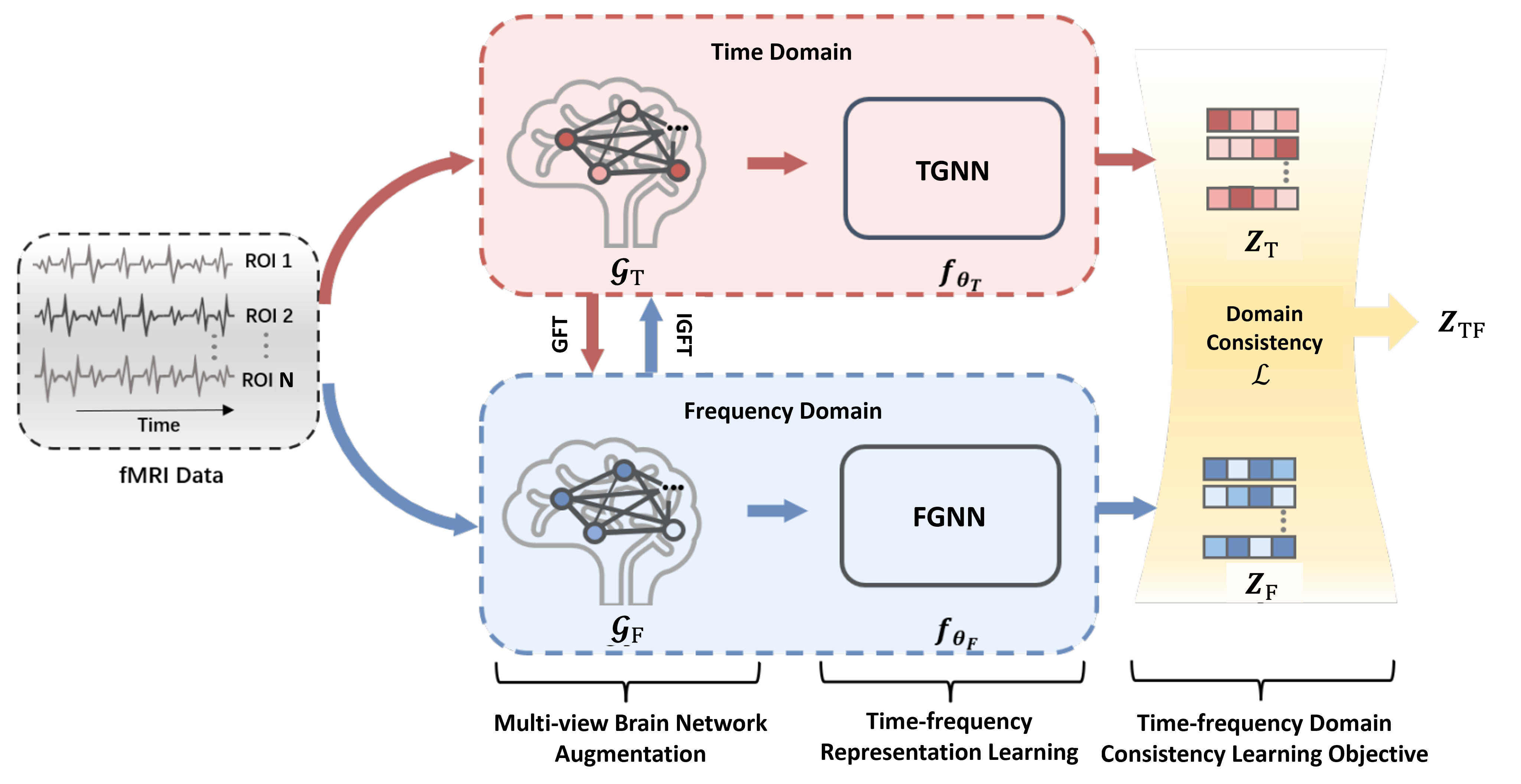} % 确保文件名和路径正确
        \caption{Overview of FENet model. The FENet model consists of three modules: multi-view graph construction, time-frequency representation learning, and learning objective.}
        \label{fig: framework}
    \end{figure*}

    The overall framework of the proposed FENet model is illustrated in Figure~\ref{fig: framework}. The core idea can be summarized as follows:(i) Constructing biologically meaningful brain networks from fMRI data in both the time and frequency domains; (ii) Employing domain-specific encoders to exploit the time-frequency characteristics in multi-view brain graphs fully; (iii) Leveraging a domain consistency objective to integrate time- and frequency-domain representations. 
	
	\subsection{Problem Definition}\label{sec3.1}
	
	Let $\{\mathbf{X}^1, \mathbf{X}^2 \dots, \mathbf{X}^ M\}$ represent $M$ samples of fMRI data and $\{ y^1, y^2, \dots, y^ M \}$, $y^ m\in\{0,1\}$ represent the corresponding labels. 
	Each of $\mathbf{X}^m=[\mathbf{x}^{1,m}, \dots,\mathbf{x}^{N,m}] \in \mathbb{R}^{ N\times D}$ represents the BOLD signals for $N$ ROIs in brain networks and $D$ is the dimension of each signal.
	Given $\mathbf{X}$, the main objective of this study is to find the representation of $\textbf{Z}_{\text{TF}}$, which effectively integrates information from both the time and frequency domains. Specifically, $\textbf{Z}_{\text{TF}}$ is obtained by training a mapping function $f_{\theta_{\text{T}}, \theta_{\text{F}}}$,
    where $\theta_{\text{T}}$ and $\theta_{\text{F}}$ denote the parameters of the time-domain and frequency-domain encoders, respectively. These encoders transform the input fMRI data into meaningful representations for downstream tasks. To this end, we formulate the task as a graph classification problem, aiming to optimize the mapping function such that $f_{\theta_{\text{T}}, \theta_{\text{F}}}(\mathbf{X}) \rightarrow y$.

	\subsection{Multi-view Brain Network Augmentation}\label{sec3.2}

    \textcolor{black}{Unlike traditional augmentation methods that risk disrupting the underlying brain network structure, our approach explicitly treats the time-domain and frequency-domain representations as two augmented graph views defined on a shared graph topology but with domain-specific node features. This formulation allows us to preserve the consistent brain connectivity structure while emphasizing complementary temporal and spectral characteristics through separate encoding spaces. By doing so, we enable the self-supervised objective to explicitly align and integrate frequency information into the brain network representation, resulting in semantically consistent yet domain-aware multi-view features that improve model robustness and interpretability.}

	\begin{figure*}[t!]
        \centering
        \includegraphics[width=0.85\textwidth]{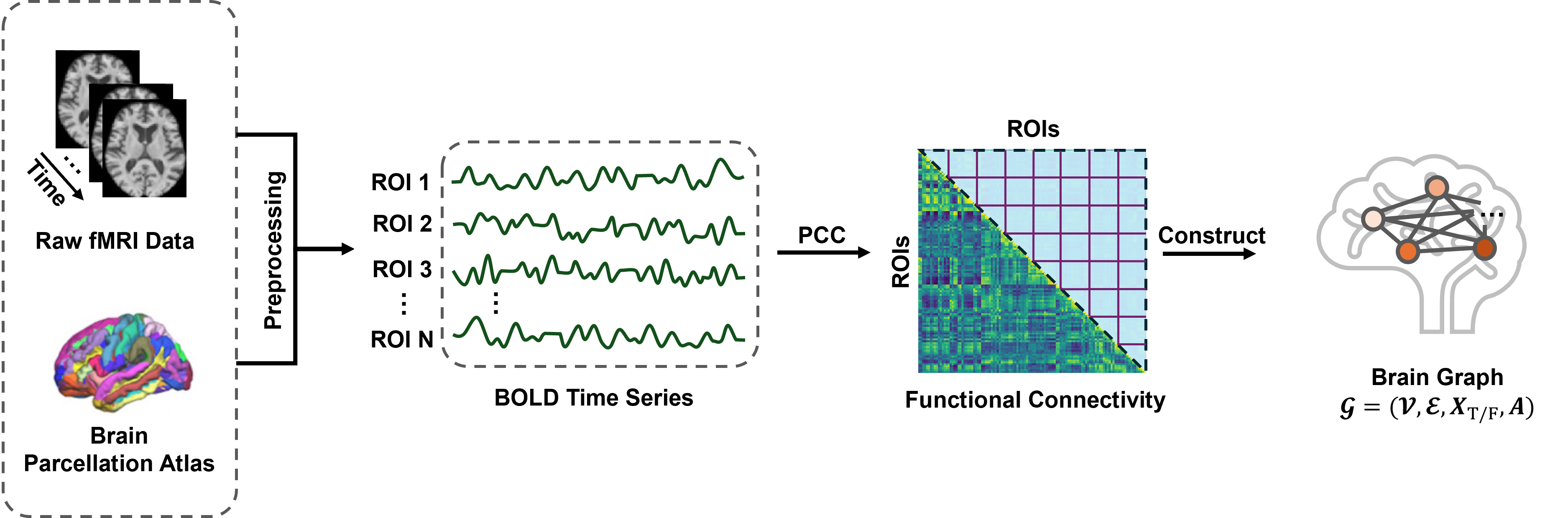} % 确保文件名和路径正确
        \caption{Pipeline of brain graph construction from raw fMRI data.}
        \label{fig: brain network}
    \end{figure*}
    
	\subsubsection{Time-domain Brain Graph Construction}

    %To ensure biological relevance, both brain network views are constructed based on functional connectivity, a widely adopted modeling approach in clinical research~\cite{du2024survey}. 
    \textcolor{black}{Following prior clinical research~\cite{du2024survey}, we represent the brain network as a graph structure. Figure~\ref{fig: brain network} shows a schematic illustration of the pipeline for constructing functional brain networks from raw fMRI data. First, the scans are preprocessed and parcellated into ROIs using a brain atlas.} The connections between pairs of ROIs are determined by functional connectivity, estimated by computing the Pearson correlation coefficient (PCC)~\cite{cohen2009pearson} between their BOLD time series data $\mathbf{X} \in \mathbb{R}^{N \times D}$, defined as:

    \begin{equation}
    a^{i,j} = \frac{\sum_{d=1}^{D} (x^{i,d} - \bar{x}^i)(x^{j,d} - \bar{x}^j)}
    {\sqrt{\sum_{d=1}^{D} (x^{i,d} - \bar{x}^i)^2} \sqrt{\sum_{d=1}^{D} (x^{j,d} - \bar{x}^j)^2}},
    \label{eq:PCC}
    \end{equation}
where $x^{i,d}$ and $x^{j,d}$ denote the BOLD signals of the $i$-th and $j$-th ROI at time point $d$, respectively, and $\bar{x}^i$ and $\bar{x}^j$ are their mean values over time. Based on the resulting connectivity strengths $a^{i,j}$, we construct the brain graph $\mathcal{G} = (\mathcal{V}, \mathcal{E}, \mathbf{X}_\text{T/F}, \mathbf{A})$. Here, the node set $\mathcal{V}=\{v^1,v^2, \dots, v^N\}$ consists of $N$ ROIs; the feature matrix $\mathbf{X}_\text{T/F}$ is defined in the augmented graph data domain; the adjacency matrix $\mathbf{A}\in \mathbb{R}^{N \times N}$ stores the functional connectivity strengths. \textcolor{black}{To construct a biologically meaningful and interpretable brain network structure while reducing noise, we retain the top $20\%$ of positive correlation values to define the edge set $\mathcal{E}$, following prior work in sparse graph construction for functional connectivity~\cite{rubinov2010complex, fornito2013graph}. Negative correlations are excluded due to their unclear physiological interpretability and potential confounding factors~\cite{murphy2017towards}.} For the brain network in the time domain, the characteristic of each node is represented by its corresponding BOLD signal, resulting in the representation of the graph $\mathcal{G}_{\text{T}} = (\mathcal{V}, \mathcal{E}, \mathbf{X}_{\text{T}}, \mathbf{A})$, where $\mathbf{X}_{\text{T}} \in \mathbb{R}^{N \times D}$.

	\subsubsection{Frequency-domain Brain Graph Construction}

    Unlike the traditional Fourier Transform performed on time series data, our approach leverages the GFT on graph structure data to generate frequency-domain brain networks. This enables the extraction of spectral representations and the identification of intrinsic neural oscillations that align with brain connectivity patterns. 
    Specifically, given a time-domain brain graph  $\mathcal{G}_{\text{T}}^{ }=(\mathcal{V},\mathcal{E},\mathbf{X}_{\text{T}},\mathbf{A})$, we transform the brain signals $\mathbf{X}_{\text{T}}$ from the time domain to the spectral domain ${\mathbf{X}_{\text{F}}}$ using the GFT operation $\mathcal{F}_G(\cdot)$: \({\mathbf{X}_{\text{F}}} = \mathcal{F}_G(\mathbf{X}_{\text{T}})=\mathbf{U}^{\top} \mathbf{X}_{\text{T}}\). 
    This formulation allows us to explicitly introduce frequency information in the model and analyze brain activity patterns from a spectral perspective as $\mathcal{G}_{\text{F}}=(\mathcal{V}, \mathcal{E},\mathbf{X}^{ }_{\text{F}},\mathbf{A}^{ })$, where $\mathcal{V}, \mathcal{E}$, and $\mathbf{A}^{ }$ are defined the same as its time-domain brain graph. Here, $\mathbf{X}^{ }_{\text{F}}\in \mathbb{R}^{N \times D}$ is the transformed feature matrix in the graph spectral domain, where each row of $\mathbf{X}^{ }_{\text{F}}$ represents a spectral coefficient corresponding to an eigenvalue $\lambda_i$. 
	
	\subsection{Time-frequency Representation Learning}\label{sec3.3}
    Diverging from conventional SSL models that focus solely on encoding two graph views within the time domain, our approach employs distinct graph encoders for multi-view brain networks. Specifically, we define the TGNN encoder $f_{\theta_{\text{T}}}(\cdot)$ for the time domain and FGNN encoder $f_{\theta_{\text{F}}}(\cdot)$ for the frequency domain. 
     This tailored design facilitates more effective encoding of the complementary information inherent in fMRI data.

     \begin{figure*}[t!]
        \centering
        \includegraphics[width=0.7\textwidth]{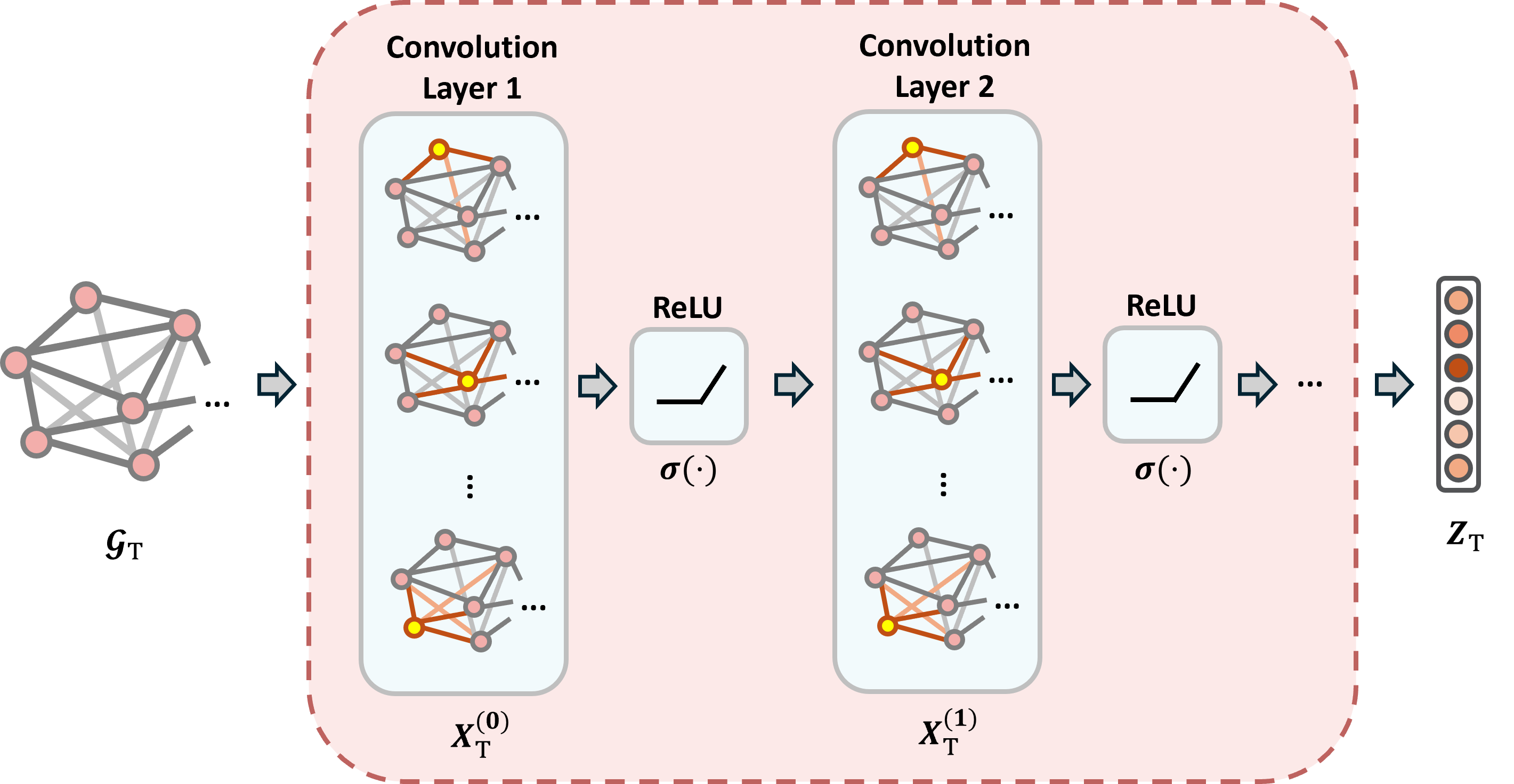} % 确保文件名和路径正确
        \caption{\textcolor{black}{The framework of the TGNN module consists of several GCN layers.}}
        \label{fig: TGNN}
    \end{figure*}
    
	\subsubsection{Time-domain Graph Neural Network Encoder}
    Graph convolutional networks (GCNs) \cite{zhang2019graph,9416834} are employed as the TGNN encoder due to their effectiveness in handling non-Euclidean and irregular data structures, which are characteristic of fMRI data, as shown in Figure~\ref{fig: TGNN}. Specifically, we leverage GCNs to model the interdependencies between BOLD signals across ROIs, enabling time-domain feature extraction by aggregating information from neighboring ROIs.  
	More specifically, we obtain the node features $\textbf{X}_{\text{T}}^{(l+1)}$ at the $(l+1)$-th convolution layer from the preceding layer as:
    
	\begin{equation}
		\label{GCN}
		\textbf{X}_{\text{T}}^{(l+1)}=\sigma\left(\tilde{\textbf{D}}^{-1/2}\tilde{\textbf{A}^{ }} \tilde{\textbf{D}}^{-1/2} \textbf{X}_{\text{T}}^{(l)}\textbf{W}^{(l)}\right),
	\end{equation}
	where %$\textbf{X}_{\text{T}}^{(l+1)}$ represents the updated node features in the $(l+1)$-th hidden layer, 
    $\tilde{\textbf{A}^{ }}=\mathbf{A}+\mathbf{I}$ is the adjacency matrix of the graph with added self-loops, $\tilde{\textbf{D}}$ is the diagonal degree matrix of $\tilde{\textbf{A}}$, $\mathbf{W}\in \mathbb{R}^{N \times N}$ is the learnable parameter matrix, $\sigma(\cdot)$ denotes the activation function (i.e., Rectified Linear Unit (ReLU) function), and $\textbf{X}_{\text{T}}^{(l)}$ denotes the node features in $l$-th hidden layer, where it is noteworthy that the node feature in the $0$-th layer hidden layer is the input itself. Finally, we consider the node features learned in the last layer of graph convolution as the time-domain representation, which denotes as $\mathbf{Z}_{\text{T}}^{ } = f_{\theta_{\text{T}}}(\mathcal{G}_{\text{T}}), \mathbf{Z}_{\text{T}} \in \mathbb{R}^{ N \times D}$.  

   \begin{figure*}[t!]
        \centering
        \includegraphics[width=0.9\textwidth]{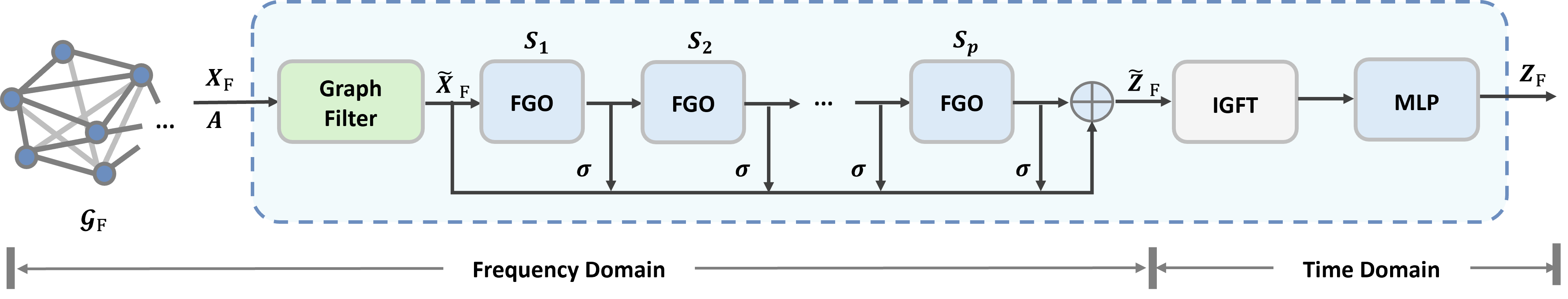} % 确保文件名和路径正确
        \caption{The framework of the FGNN module consists of a graph filter and several FGO network layers.}
        \label{fig: FGNN}
    \end{figure*} 
	\subsubsection{Frequency-domain Graph Neural Network Encoder}
 
    To efficiently capture disease-related frequency information with acceptable computational complexity, we introduce a novel frequency-domain encoder, FGNN, as shown in Figure~\ref{fig: FGNN}. 
    This module consists of two primary components: a graph filter and FGO layers. The graph filter captures the crucial frequency components by integrating biological-inspired prior knowledge. Meanwhile, the FGO network introduces log-linear multiplication to replace standard linear multiplication, improving model scalability.% while  
	
	\noindent \textbf{Graph Filter.} 
    Spectral graph theory provides a robust framework for analyzing the structural properties of brain networks, where the spectral components naturally correspond to frequency characteristics, offering insights into both global and local connectivity patterns. Specifically, the GFT decomposes BOLD signals into different graph frequency components. Low-frequency components, associated with smaller Laplacian eigenvalues, capture globally coherent brain activity, reflecting stable functional connectivity across the brain network. In contrast, high-frequency components, corresponding to larger eigenvalues, emphasize localized variations and abrupt changes, often linked to abnormal or disrupted brain regions—patterns commonly observed in psychiatric disorders~\cite{huang2016graph}. This spectral decomposition enables in-depth frequency-based analysis of brain networks, facilitating both the examination of frequency-specific information and the extraction of informative features for downstream tasks. 
    
    In this study, we leverage the eigenvalues $\mathbf{\Lambda}$ of the brain networks to decompose brain frequency components into distinct bands: low-frequency components within $\lambda_\text{min}$ and $\lambda_\text{L}$, high-frequency components within $\lambda_\text{H}$ and $\lambda_\text{max}$, and the remaining range considered as middle-frequency components. 
    To examine how different frequency components relate to cognitive states or mental disorders, we design corresponding graph filters that enhance learning at each specific frequency range, capturing more biologically relevant information and improving model interpretability. Specifically, a low-pass graph filter \(\mathbf{H}_{\text{L}} = \text{diag}({h}_{\text{L}}), \mathbf{H}_{\text{L}}\in \mathbb{R}^{N \times N}\) is defined, where \({h}_{\text{L}}\) takes the value of 1 for eigenvalues \((\lambda_{\text{min}}<\lambda_i<\lambda_{\text{L}})\) and 0 otherwise. Similarly, a band-pass filter \(\mathbf{H}_{\text{M}}\in \mathbb{R}^{N \times N}\) and a high-pass filter \(\mathbf{H}_{\text{H}}\in \mathbb{R}^{N \times N}\) are constructed based on their respective frequency bands. 
    After filtering, we obtain the filtered signals as \(\mathbf{X}_{\text{FH}}\textquotesingle =\mathbf{H}_{\text{H}}\mathbf{X}_{\text{F}}\), \(\mathbf{X}_{\text{FM}}\textquotesingle =\mathbf{H}_{\text{M}}\mathbf{X}_{\text{F}}\), and \(\mathbf{X}_{\text{FL}}\textquotesingle =\mathbf{H}_{\text{L}}\mathbf{X}_{\text{F}}\), corresponding to the high-, middle-, and low-frequency components of the brain network, respectively. \textcolor{black}{Then, among these filtered spectral components, we selectively integrate those that capture more significant information to construct the final filtered feature matrix $\tilde{\mathbf{X}}_\text{F} \in \mathbb{R}^{N \times K}$, where $K$ denotes the dimension of the selected frequency components. This design enables the model to effectively leverage meaningful spectral characteristics of brain connectivity during representation learning.}

	\noindent \textbf{Fourier Graph Operator.} 
	Encoding frequency-domain features on graphs often incurs high computational costs due to complex eigendecomposition and graph convolution operations. To alleviate this burden, FGNN incorporates an efficient learnable layer, the FGO, inspired by FourierGNN~\cite{yi2023fouriergnn}, significantly reducing computational overhead while effectively capturing frequency information. Specifically, FGNN extracts frequency information by using $N$-invariant FGOs, denoted as $\mathbf{S} \in \mathbb{R}^{ N \times K \times K}$, enabling straightforward matrix multiplication within the frequency-domain brain networks, as shown in Figure~\ref{fig: FGNN}. Here, $\mathbf{S}$ is a tailored Green\textquotesingle s kernel $\kappa$ \cite{yi2023fouriergnn}, a translation-invariant kernel where $\kappa[i,j] = \kappa[i-j]$ represents its invariance property when sliding from node $v^i$ to node $v^j$.   
    This operation transforms the $([N] \times [N])$ matrix into a $\mathbb{R}^{K \times K}$ matrix-valued projection. The FGO parameterizes graph convolutions analogous to the shared-weight convolution kernels in traditional convolutional neural networks (CNNs). The operation is defined as: $\mathcal{F}_G^{-1}(\mathbf{X}_{\text{F}}\mathbf{S}^{0:p})$, which equals to $\mathbf{A}^{p:0}\mathbf{X}_{\text{T}}\mathbf{W}^{0:p}$ in time domain. Here, $\mathcal{F}_G^{-1}(\cdot)$ is the IGFT, $\mathbf{A}^p \in \mathbb{R}^{N \times N} $ represents as the $p$-th sparsity pattern of $\mathbf{A}$, $\mathbf{W}^p \in \mathbb{R}^{K \times K}$ is the $p$-th weight matrix. This convolutional kernel operates over the input features through matrix multiplications of size $([K] \times [K])$ in the Fourier space. By stacking multiple FGO layers, the model can efficiently capture informative features while maintaining log-linear computational complexity. 
	Thus, FGNN performs a recursive multiplication between the filtered frequency-domain brain graph embeddings $\tilde{\mathbf{X}}_{\text{F}}$ and FGO layers $\mathbf{S}^{0:p}$ to obtain the representation $\tilde{\mathbf{Z}}_{\text{F}}^{ } \in \mathbb{R}^{N \times K}$ in Fourier space. This process can be defined as:
	\begin{equation}
		\tilde{\mathbf{Z}}_{\text{F}} = \sum_{p=0}^P \sigma\left(\tilde{\mathbf{X}}_{\text{F}}^{ } \mathbf{S}^{0:p}+b^p\right),     \mathbf{S}^{0:p} = \prod_{j=0}^{p} \mathbf{S}^j,
	\end{equation}
	where $p$ is the number of FGO layers, $\mathbf{S}^j \in \mathbb{R}^{K \times K}$ is the $p$-th step FGO, $b^p \in \mathbb{R}^{K}$ are the complex-valued biases parameters.

    Our model aims to minimize the distance between the time- and frequency-domain representations in the latent time-frequency space, thereby learning domain-invariant features. This objective is further discussed in Section~\ref{sec3.4}. To achieve this, we transform the learned frequency-domain representations $\tilde{\mathbf{Z}}_{\text{F}}^{ }$ in Fourier space back into the time-space, facilitating the alignment of representations from both domains and contributing to more effective model training.
    This conversion is achieved by applying IGFT operation $\mathcal{F}_G^{-1}(\cdot)$ as follows: $\mathcal{F}^{-1}_G(\tilde{\mathbf{X}}_{\text{F}})= \mathbf{U}\tilde{\mathbf{X}}_{\text{F}}$. The resulting representation is then passed through a multilayer perceptron (MLP) to project it into the time-frequency latent space, ensuring alignment with the representation learned in the time domain. Finally, we denote the representation learned from the frequency domain as: $\mathbf{Z}_{\text{F}}^{ } = f_{\theta_{\text{F}}}(\mathcal{G}_{\text{F}}), \mathbf{Z}_{\text{F}}^{ } \in \mathbb{R}^{ N \times D}$.

    \subsubsection{Computation Complexity Analysis}
    This section highlights the improved efficiency of encoding frequency-domain features in SSL settings due to the use of our FGNN. 
    Assume the frequency-domain graphs consist of $N$ nodes and $|\mathcal{E}|$ edges, and let each node have $D$ dimension features. Traditional methods apply GCN that performs matrix multiplications for aggregating node features. This leads to a computational complexity of $O(2N|\mathcal{E}|^2 + 2ND^2)$. In contrast, our proposed FGNN replaces conventional GCN layers with an efficient feature encoding process involving graph filtering and recursive multiplications utilizing the efficient FGO layers. We suppose $K$ frequency components represent the filtered node features in the Fourier space, where $K \ll N$ and the FGO operator has size $K \times K$. We apply a third-order summation for the FGOs, yielding a total computational complexity for the frequency-domain encoder of $O(NK log N + 3NK^2)$, including GFT and IGFT operations. Overall, the log-linear complexity significantly reduces computational overhead while preserving essential frequency-domain information.

    \begin{figure}[t!]
        \centering
        \includegraphics[width=0.45\textwidth]{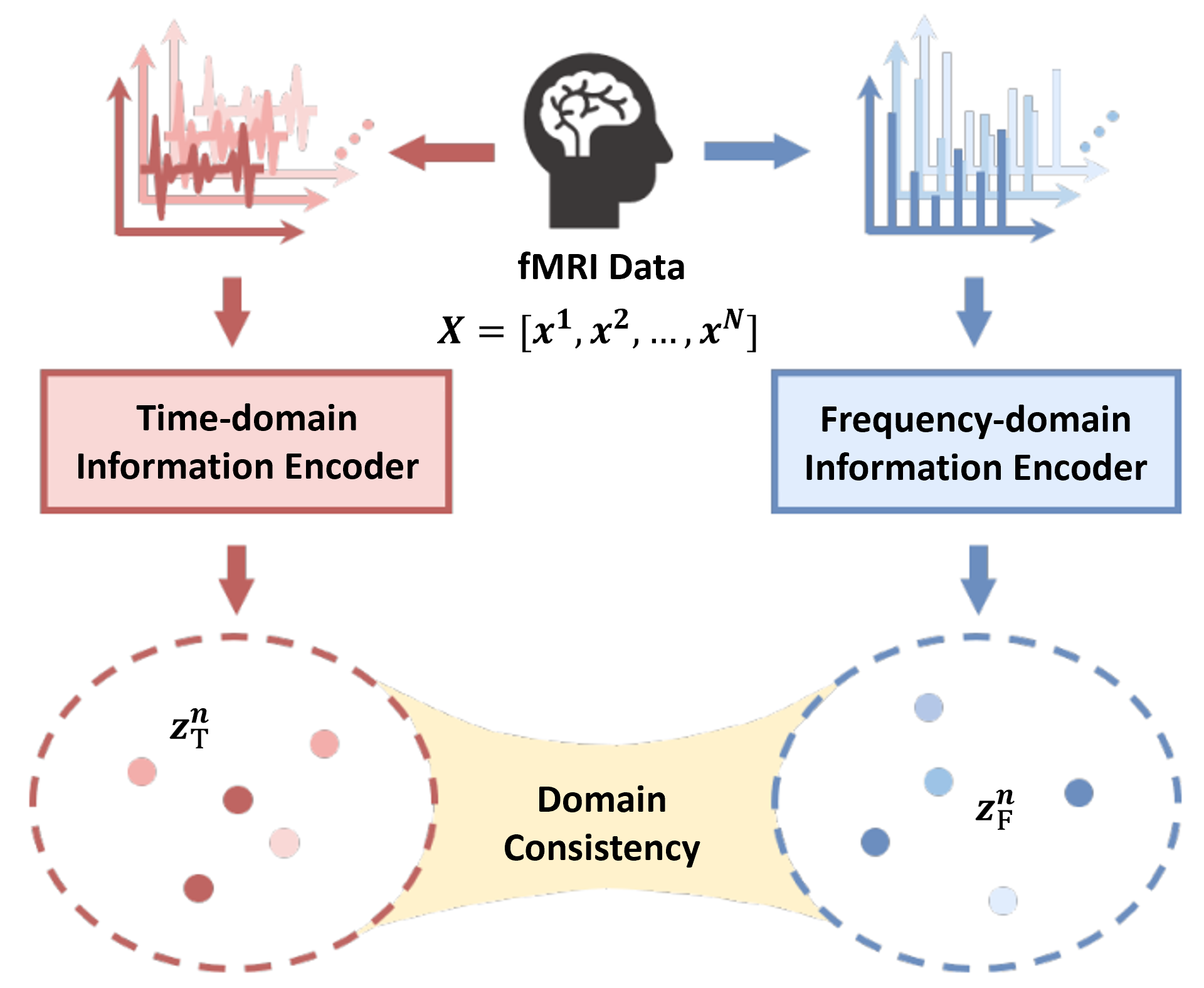} % 确保文件名和路径正确
        \caption{Illustration of the learning objective. For each BOLD signal $\textbf{x}^ n$, its time-domain representation $\textbf{z}_{\text{T}}^{ n}$ and the frequency-domain representation $\textbf{z}_{\text{F}}^{ n}$ are obtained via their respective domain-specific encoders. The learning objective is to ensure that these two representations remain close in the latent time-frequency space, promoting consistency between the two domains.}
        \label{fig: TF-C}
    \end{figure}

	\subsection{Time-Frequency Domain Consistency Learning Objective}\label{sec3.4}
    Our learning objective employs a non-contrastive SSL approach for graph-level representation learning, eliminating the need for contrastive pairs. Guided by the concept of time-frequency domain consistency (as illustrated in Figure~\ref{fig: TF-C}), the objective enforces proximity between the time- and frequency-domain representations of the same BOLD signal in the latent space. This encourages the model to capture domain-invariant features, facilitating the comprehensive integration of diverse information from fMRI data. Specifically, we extend the application of CCA to the SSL framework. The correlation term aims to maximize the alignment between the two graph views by minimizing the distance between their representations in the latent time-frequency space, thereby extracting domain-invariant features. Additionally, we introduce a regularization term (decorrelation term) for each domain to prevent model collapse. This term encourages the learned representations to retain domain-specific semantic information by maintaining diversity across different feature dimensions. The overall process, taking the time-domain and frequency-domain representations as input, can be formulated as:
    
	%\begin{equation}
	%	\label{Loss}
	%	\mathcal{L} = \frac{1}{M} \sum_{m=1}^ m \frac{\mathbf{z}_{\text{T}}^{ n} \cdot \mathbf{z}_{\text{F}}^{ n}}{\|\mathbf{z}_{\text{T}}^{ n}\| \|\mathbf{z}_{\text{F}}^{ n}\|} + \lambda \left\| (\mathbf{Z}_{\text{T}}^{ })^{\top}\mathbf{Z}_{\text{T}}^{ } - \mathbf{I} \right\|_{\text{F}}^2 + \gamma \left\| (\mathbf{Z}_{\text{F}}^{ })^{\top}\mathbf{Z}_{\text{F}} - \mathbf{I} \right\|_{\text{F}}^2,
	%\end{equation}
	\begin{equation}
		\small
		\label{Loss}
		\mathcal{L} =  \left\| \mathbf{Z}_{\text{T}}^{ }- \mathbf{Z}_{\text{F}} \right\|_{\text{F}}^2 +   \gamma \left\| \mathbf{Z}_{\text{T}}^{\top}\mathbf{Z}_{\text{T}} - \mathbf{I} \right\|_{\text{F}}^2 + \beta \left\| \mathbf{Z}_{\text{F}}^{\top}\mathbf{Z}_{\text{F}} - \mathbf{I} \right\|_{\text{F}}^2,
	\end{equation}
	%By minimizing the loss function, we aim to maximize the similarity between time-domain and frequency-domain representations in the latent space while capturing their respective semantically rich features. 
    where $\| \cdot \|_\text{F}$ denotes the Frobenius norm, and $\gamma$ and $\beta$ are trade-off coefficients. The first term minimizes the disparity between the two graph views, enhancing their alignment. The second and third terms serve as regularization losses, preserving the semantic features of each domain. 
    By minimizing the loss function, the model learns optimized representations from both domains. The final fused representation $\mathbf{Z}_{\text{TF}}$ is then obtained by averaging these optimized time- and frequency-domain representations, effectively integrating information from both domains.
    %The enhanced brain graph representation is obtained and ultimately used for disease detection by minimizing the loss function. \td{? half sentence?}%$\mathbf{Z}_{\text{TF}}$\td{how do you get ztf, which is not mentioned anywhere?}
 
    The overall training process of FENet is outlined in Algorithm \ref{algorithm}. Initially, we apply the encoders for each domain graph separately to learn enhanced brain signal representations by optimizing Equation \eqref{Loss}. After convergence during the pretraining, a subset of data is used for model fine-tuning, further refining the representation learning process.
    
	\begin{algorithm}
        \caption{Overview of FENet Training}
        \label{algorithm}
        \begin{algorithmic}[1]
            \State \textbf{Input:} The fMRI datasets for $M$ samples and their labels $\{\mathbf{X}^{i}, y^{i}\}_{i=1}^ M$; Training epoch $E$.
            \State \textbf{Output:} Trained encoders $f_{\theta_{\text{T}}}(\cdot)$ and $f_{\theta_{\text{F}}}(\cdot)$.
            \State \textbf{Initialize:} The model parameters $\theta_{\text{F}}$ and $\theta_{\text{F}}$.
            \State \textit{Pretraining:}
            \For{$i=1,2,\dots, M$}
            \State \textbf{Compute} multi-view brain graphs $\mathcal{G}_{\text{T}}^{i}$ and $\mathcal{G}_{\text{F}}^{i}$ in time and frequency domain.
                \For{$t = 1, 2, \dots, E$}
                    \State \textbf{Compute} $\mathbf{Z}_{\text{T}}^{i} = f_{\theta_{\text{T}}}(\mathcal{G}_{\text{T}}^{i}) \triangleright$ Obtain time-domain representations
                    \State \textbf{Compute} $\mathbf{Z}_{\text{F}}^{i} = f_{\theta_{\text{F}}}(\mathcal{G}_{\text{F}}^{i}) \triangleright$ Obtain frequency-domain representations
                    \State \textbf{Optimize} model parameters $\theta_{\text{F}}$ and $\theta_{\text{F}}$ with loss function (see equation~\ref{Loss}).
                \EndFor 
            \EndFor
            \State \Return Brain graph representations $\{\mathbf{Z}_{\text{TF}}^{i}\}_{i=1}^M$
            \State \textit{Fine-tunning:}
                \For{$t = 1, 2, \dots, E$}
                \For{Each batch in $\{\mathbf{Z}_{\text{TF}}^i,y^{i}\}_{i=1}^ M$}
                    \State \textbf{Compute} $\hat{y}^i = \phi(\mathbf{Z}_{\text{TF}}^i) \triangleright$ Predicted labels with classifier
                    \State \textbf{Optimize} $\phi(\cdot)$ with $\mathcal{L}_{class}(\hat{y}^i,{y}^i)$.
                \EndFor 
            \EndFor 
        \end{algorithmic}
    \end{algorithm}

\section{Experiments}
\label{Experiments}
    \subsection{Datasets}
    We conducted our experiments on two real-world medical datasets: the Autism Brain Imaging Data Exchange (ABIDE)~\footnote[1]{\url{https://fcon_1000.projects.nitrc.org/indi/abide/}} and the Attention-Deficit Hyperactivity Disorder (ADHD-200)~\footnote[2]{\url{https://fcon_1000.projects.nitrc.org/indi/adhd200/}} datasets. 

    \noindent\textbf{ABIDE Dataset.} This database aggregates data from 17 different acquisition sites, offering resting-state fMRI scans from 1,112 participants, including individuals diagnosed with ASD and healthy controls (HC). Additionally, it includes phenotypic information on executive functioning, language, visuospatial ability, motor functioning, and emotional status, all assessed via self-report. \textcolor{black}{For this study, we carefully curated the dataset by excluding scans with incomplete imaging data, missing phenotypic information, or time series of insufficient length to ensure both high data quality and the ability to capture temporal characteristics. The final sample included 743 participants, 364 with ASD (aged 7–64) and 379 HC (aged 6–42), providing a relatively balanced and demographically matched cohort for analysis.}

    \noindent\textbf{ADHD-200 Dataset.} Collected from eight independent imaging sites, the ADHD-200 dataset was designed to facilitate early diagnosis of ADHD and inform clinical treatment decisions. It includes detailed phenotypic data such as diagnostic status, dimensional ADHD symptom measures, age, sex, intelligence quotient (IQ), and lifetime medication status. For our experiments, we employed a sample of 459 participants, comprising 229 typically developing individuals and 230 children and adolescents with ADHD (aged 7-21).

    \subsection{Data Preprocessing}\label{Appendix_preprocessing}

    In this study, we obtained fMRI data from both the ABIDE and ADHD datasets. We then processed these fMRI signals using the graph theoretical network analysis (GRETNA) toolbox\footnote[3]{\url{https://www.nitrc.org/projects/gretna/}.}, which operates in conjunction with the SPM12 software\footnote[4]{\url{https://www.fil.ion.ucl.ac.uk/spm/software/spm12/}.}. The preprocessing steps included slice timing correction, head motion correction, spatial normalization, and Gaussian smoothing. Next, the automated anatomical labeling (AAL) atlas was used as the reference space to divide the brain into 116 ROIs, from which BOLD time series were subsequently extracted. \textcolor{black}{To ensure consistency across subjects and datasets, all BOLD time series were uniformly truncated or padded to retain 170 time points.}

    \subsection{Baselines}
    For baseline methods used in the comparison, we included the GNN-based approaches, BrainGNN~\cite{li2021braingnn} and BrainGB~\cite{cui2022braingb}, which employ supervised learning strategies. Additionally, we considered three contrast-based SSL methods, GCA~\cite{zhu2021graph}, SpCo~\cite{liu2022revisiting}, and MA-GCL~\cite{gong2023ma}, along with two similarity-based SSL methods, GATE~\cite{peng2022gate} and CCA-SSG~\cite{zhang2021canonical}. Open-source codes for all comparison methods are adopted, and the grid search technology is performed to determine their optimal hyperparameters.
	To ensure a fair comparison, we apply the same dynamic window strategy and utilize a grid search technique to identify the best practices for each comparison method. Details for each comparison method are outlined below.
	\begin{itemize}
		\item BrainGNN~\footnote[5]{\url{https://github.com/xxlya/BrainGNN_Pytorch}}~\cite{li2021braingnn} designs a novel ROI-aware GCN model combined with ROI-selection pooling layers to improve the interpretability of analyzing fMRI data. This serves as the baseline graph-based method for fMRI data.
		\item BrainGB~\footnote[6]{\url{https://github.com/HennyJie/BrainGB}}~\cite{cui2022braingb} comprehensively summarizes the current methods for constructing brain networks and proposes a generic GNN architecture to standardize the process of brain network analysis. %Serving as a benchmark for brain network analysis, BrainGB is designed to provide a standardized foundation for research. 
		\item GCA~\footnote[7]{\url{https://github.com/CRIPAC-DIG/GCA}}~\cite{zhu2021graph} introduces a novel adaptive enhancement method for contrast-based SSL. This approach perturbs the topology and semantics of graphs as priors, highlighting important structural connections while preserving crucial semantic information at the node level.
		\item SpCo~\footnote[8]{\url{https://github.com/liun-online/SpCo}}~\cite{liu2022revisiting} combines graph spectral information with SSL, addressing imbalances in spectral learning. The method introduces a spectral-contrastive learning module that effectively enhances existing SSL methods. 
		\item 
		MA-GCL~\footnote[9]{\url{https://github.com/GXM1141/MA-GCL}}~\cite{gong2023ma} designs different encoders to enhance view diversity, addressing the issue of semantic label changes caused by graph perturbations. It employs three techniques to reduce high-frequency noise and enhance view diversity safely.
		\item CCA-SSG~\footnote[10]{\url{https://github.com/hengruizhang98/CCA-SSG}}~\cite{zhang2021canonical} proposes an efficient similarity-based SSL framework by utilizing the CCA as the learning objective. This allows the structure of this model to be concise and achieves significant results compared with other SSL methods. 
		\item GATE~\footnote[11]{\url{https://github.com/LarryUESTC/GATE}}~\cite{peng2022gate} presents a similarity-based SSL method, introducing the concept of a population graph and utilizing an enhanced CCA method to improve the sensitivity of model to spurious signals. 
        \item \textcolor{black}{BraGCL}~\footnote[12]{\url{https://github.com/XuexiongLuoMQ/BraGCL-framework}}~\cite{luo2024interpretable} \textcolor{black}{proposes an interpretable contrast-based SSL method that constructs augmented views by perturbing less important edges and attributes, thereby learning discriminative patterns from hard negative pairs.}
        \item \textcolor{black}{A-GCL}~\footnote[13]{\url{https://github.com/qbmizsj/A-GCL}}~\cite{zhang2023gcl} \textcolor{black}{proposes an adversarial graph contrastive learning framework that generates positive and negative sample pairs using a Bernoulli mask, enabling the model to learn label-independent graph embeddings.}
	\end{itemize}
	\textcolor{black}{For supervised learning baselines such as BrainGNN and BrainGB, we follow standard practice by adopting a consistent data split of 80\% training, 10\% validation, and 10\% testing across all experiments, ensuring full utilization of available labeled data. } All SSL baselines are implemented based on their original publications and official code. For contrastive SSL methods (i.e., GCA, SpCo, MA-GCL, BraGCL, and A-GCL), positive and negative node sample pairs are constructed by applying corruption functions to the input graph. In contrast, similarity-based SSL methods (i.e., CCA-SSG and GATE) generate two augmented graph views through a random masking strategy. \textcolor{black}{To ensure a fair comparison, we conduct hyperparameter tuning for each baseline on our datasets, selecting optimal configurations based on performance on the validation set.}

    \subsection{Implementation Details}
    
    We implement FENet in PyTorch and conduct all experiments on an NVIDIA Tesla P100 GPU with 16GB of memory. Model parameters are optimized using the AdamW optimizer~\cite{loshchilov2017decoupled} with a learning rate of $1e^{-5}$. The time-domain information encoder is configured as a standard two-layer GCN model to mitigate the over-smoothing issue in GCN. Based on empirical evaluations, the frequency-domain information encoder comprises three FGO layers for effective feature extraction in the frequency-domain graph. To effectively set the frequency thresholds for low-pass, band-pass, and high-pass filters in graph signal processing, we optimized the $\lambda_\text{L}$ as the lowest $20\%$ of eigenvalues, while the $\lambda_\text{H}$ as the top $20\%$ of eigenvalues. 
    To evaluate the performance of both encoders, we tune the hyperparameters in Equation~\eqref{Loss} and compare the model classification accuracy; ultimately, $\gamma$ and $\beta$ are set to $1e^{-5}$ and $1e^{-4}$. \textcolor{black}{The training procedure comprises two distinct phases: self-supervised pre-training and supervised fine-tuning. During the pre-training phase, the model is trained in a fully unsupervised manner for 200 iterations on the entire dataset without using any label information. In the fine-tuning phase, we randomly select 20\% of the labeled samples from the training split to train the classifier head.} For robust evaluation, we employ 5-fold cross-validation, ensuring that the data is split into training, validation, and test sets in a stratified manner for each fold. We repeat the entire process five times with different random seeds to assess variability. The final reported results include the mean and standard deviation across these runs. We evaluate classification performance using accuracy, area under the ROC curve (AUC), recall, and F1-score, reporting both the mean and standard deviation for each metric.
        
\section{Results and Discussion}     
\label{Results and Discussion}
    
    \subsection{Performance Comparison}

    We evaluated the proposed model against a range of strong baseline methods, as shown in Table \ref{Performance comparison}. Notably, some existing SSL approaches achieve comparable or even superior performance to fully supervised methods, highlighting the effectiveness of SSL frameworks in learning meaningful brain graph representations under limited labeled data. However, the performance of SSL methods is not uniform. Methods such as MA-GCL and GCA exhibit relatively lower performance, which can be attributed to their reliance on manually designed contrastive objectives. These objectives are not well-suited for distinguishing HC from patients, as the classification depends on global brain graph structures rather than individual node features. \textcolor{black}{In contrast, brain-network-specific methods like A-GCL and BraGCL achieve better performance but rely on perturbation-based data augmentation strategies that can inevitably distort brain network structures or introduce noise.} On the other hand, similarity-based methods provide greater flexibility and robustness by avoiding explicit contrastive pairs, but they often overlook critical frequency-domain information necessary for capturing neural oscillatory patterns associated with psychiatric disorders. Although SpCo includes high-frequency spectral cues during graph perturbations, its analysis remains confined to time-domain representations, limiting its ability to extract and leverage meaningful frequency-domain semantics.
    
    In contrast to existing approaches, our proposed FENet integrates both time-domain and frequency-domain information, enabling a more comprehensive understanding of brain activity. Experimental results clearly show that FENet outperforms all baseline methods across multiple evaluation metrics. 
    Specifically, for ASD detection, FENet improves accuracy, AUC, and F1-score by 2.7\%, 1.4\%, and 4.9\%, respectively, compared to the strongest baseline. Although the recall is slightly lower than that of A-GCL, it remains competitive. Similarly, for ADHD detection, FENet achieves gains of 3.4\% in accuracy, 0.7\% in AUC, and 2.3\% in F1-score, while the recall again falls slightly below A-GCL. 
    We attribute the relatively lower recall to the emphasis of FENet on learning highly discriminative representations through time–frequency integration and decorrelation objectives, which may lead to higher precision but occasionally miss borderline or ambiguous cases. In contrast, A-GCL relies on more aggressive graph augmentation strategies, which might promote better generalization to positive cases, thereby improving recall at the cost of precision.
    
    Nonetheless, the overall performance gains in FENet, including improvements in accuracy, F1-score, and interpretability, demonstrate the effectiveness of frequency-enhanced representation learning for capturing temporal fluctuations and neural oscillatory patterns relevant to psychiatric disorder detection. This confirms the ability of FENet to deliver robust and explainable brain graph representations.

\begin{table*}[t!]
    \centering
    \scriptsize  % 缩小字体
    \caption{Performance (\%) on the ABIDE and ADHD datasets. SL represents supervised learning, and SSL represents self-supervised learning. The best results are bold, and the second results are underlined.}
    \label{tab1}
    \setlength{\extrarowheight}{2pt} 
    \resizebox{\linewidth}{!}{  % 缩放表格以适应页面宽度
    \begin{tabular}{c|c|cccc|cccc}
    \toprule
        &\multirow{2}{*}{\textbf{Methods}}  &  \multicolumn{4}{c|}{\textbf{ABIDE}} & \multicolumn{4}{c}{\textbf{ADHD}} \\
        \cmidrule{3-10}
        && \textbf{ACC} & \textbf{AUC}  & \textbf{Recall}  & \textbf{F1-score} &  \textbf{ACC} & \textbf{AUC}  & \textbf{Recall}  & \textbf{F1-score} \\
        \midrule
        \multirow{2}{*}{\textbf{SL}} & BrainGNN  & $54.9\pm 8.7$ & $57.1 \pm 8.3$ & $55.0 \pm 6.3$ & $53.4 \pm 6.0$ & $55.8 \pm 1.2$ & $58.0 \pm 4.9$ & $46.3 \pm 4.2$ & $55.9 \pm 4.2$ \\
        &BrainGB  & $55.6\pm 4.3$ & $56.3 \pm 3.6$ & $55.4 \pm 5.2$ & $53.6 \pm 10.2$ & $56.2 \pm 7.1$ & $59.6 \pm 5.2$ & $56.2 \pm 4.6$ & ${61.6 \pm 6.3}$ \\
        \midrule
        \multirow{8}{*}{\textbf{SSL}}&MA-GCL  &$49.1\pm 7.1$ & $44.8 \pm 9.2$ & $52.5 \pm 7.6$ & $42.5 \pm 11.8$  &$50.1\pm 5.6$ & $48.5 \pm 6.2$ & $46.5 \pm 3.7$ & $45.2 \pm 6.4$ \\
        &GCA  &$52.8\pm 3.5$ & $50.6 \pm 2.4$ & $53.2 \pm 4.4$ & $52.1 \pm 2.3 $ &$50.8 \pm 2.4$ &$52.4 \pm 5.1$ &$59.4 \pm 3.6$ &$53.6 \pm 4.4$\\
        &SpCo  & $52.0\pm 3.3$ & $51.8 \pm 4.4$ & ${52.1 \pm 7.8}$ & $51.4 \pm 6.0$ & $55.8 \pm 4.1$ & $ 58.0 \pm 4.9$ &$58.5 \pm 3.7$ &$55.9 \pm 5.1$ \\
        & CCA-SSG   &  $55.0\pm 6.1$ & $57.2 \pm 6.0$ & $55.2 \pm 7.7$ & $53.3 \pm 6.7$ &$56.6 \pm 5.8$ &${60.3 \pm 2.6}$ &$60.3 \pm 4.4$ &$58.0 \pm 8.6$\\
        &GATE  & $\underline{59.8 \pm 3.4}$ & ${62.8 \pm 6.1}$ & $64.4 \pm 7.3$ & $\underline{65.0 \pm 5.2}$ &${60.1 \pm 5.2}$ &$52.6 \pm 3.8$ &${62.8 \pm 9.3}$ &$60.9 \pm 5.5$\\
        &\textcolor{black}{BraGCL} & $ \textcolor{black}{53.5\pm 4.3}$ & $ \textcolor{black}{55.1 \pm 5.2}$ & $ \textcolor{black}{53.4 \pm 4.4}$ & $ \textcolor{black}{51.8 \pm 5.0}$ & $ \textcolor{black}{63.8 \pm 5.6}$ & $ \textcolor{black}{67.3 \pm 3.7}$ & $ \textcolor{black}{63.8 \pm 5.6}$ & $ \textcolor{black}{63.7 \pm 5.5}$ \\
        &\textcolor{black}{A-GCL}   & $ \textcolor{black}{59.0\pm 1.6}$ & $ \textcolor{black}{\underline{63.3 \pm 1.8}}$ & $ \textcolor{black}{\mathbf{71.6 \pm 12.9}}$ & $ \textcolor{black}{63.0 \pm 4.1}$ & $ \textcolor{black}{\underline{64.4 \pm 2.5}}$ & $ \textcolor{black}{\underline{69.0 \pm 3.2}}$ & $ \textcolor{black}{\mathbf{68.5 \pm 17.2}}$ & $ \textcolor{black}{\underline{64.7 \pm 3.9}}$ \\
        \cmidrule{2-10}
        &\textbf{FENet} (ours)  & $\mathbf{62.5\pm 3.9}$ & $\mathbf{64.7\pm 4.1}$ & $\underline{63.3\pm 4.8}$ & $\mathbf{69.9\pm4.8}$ &$\mathbf{67.8 \pm 4.1}$ &$\mathbf{69.7 \pm 4.4}$ &$\underline{66.7 \pm 2.9}$ &$\mathbf{67.0 \pm 5.2}$\\
        \bottomrule
    \end{tabular}
    }\label{Performance comparison}
\end{table*}

    \subsection{Data-efficient Ability Study}
    \textcolor{black}{We emphasize that the ultimate success of SSL depends on its ability to effectively leverage limited labeled data, enabling models to generalize well even under sparse supervision. This property is especially critical in healthcare and neuroimaging domains, where acquiring high-quality annotations is often costly and time-consuming. Recent studies have shown that SSL can even outperform fully supervised methods in such low-label scenarios~\cite{huang2023self, huang2024systematic, peng2022gate}. }
    
\begin{figure}[t!]
        \centering
        \begin{subfigure}[b]{0.45\textwidth} % 调整宽度
            \centering
            \includegraphics[width=\textwidth]{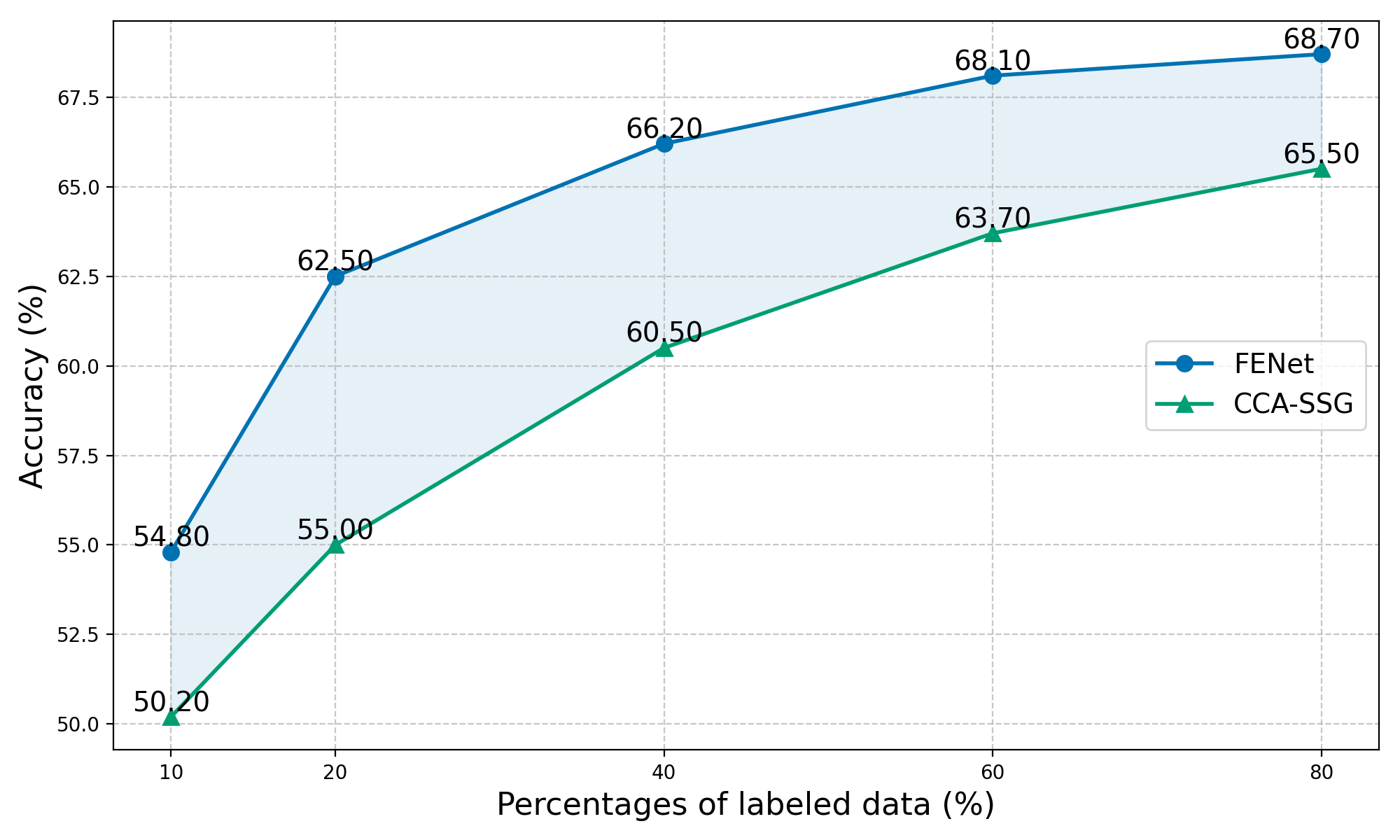}
            \caption{ABIDE}
        \end{subfigure}
        \hspace{5pt}
        \begin{subfigure}[b]{0.45\textwidth} % 调整宽度
            \centering
            \includegraphics[width=\textwidth]{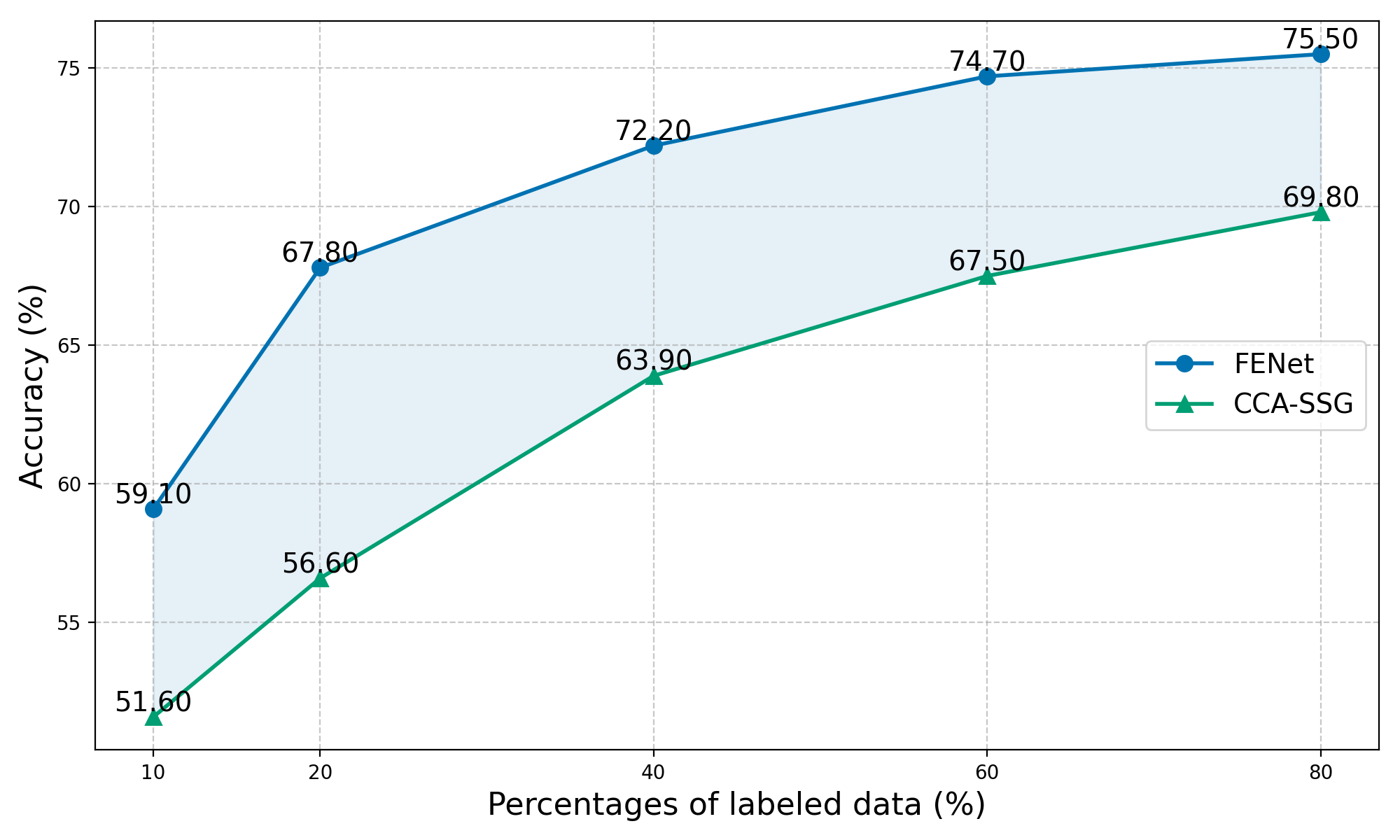}
            \caption{ADHD}
        \end{subfigure}

        \caption{Accuracy (\%) of FENet and CCA-SSG at different labeled data rates for fine-tuning.}
        \label{fig: fine-tunning}
        \end{figure}

    To evaluate the robustness and data efficiency of our proposed method under limited-supervision conditions, we fix the set of unlabeled training data and vary the proportion of labeled samples used during the fine-tuning phase. We adopt CCA-SSG as our baseline for comparison, as it also employs a CCA-based similarity-driven objective and has shown strong performance as a benchmark for non-contrastive SSL on graph data. As shown in Figure~\ref{fig: fine-tunning}, our method consistently outperforms CCA-SSG across all levels of label availability on both the ABIDE and ADHD datasets. The performance gap is particularly pronounced under extremely low-label scenarios. For instance, with only 20\% of labeled data, our model achieves 6.5\% and 11.2\% higher accuracy than CCA-SSG on the ABIDE and ADHD datasets, respectively (see Table~\ref{tab1}). Even under the most challenging setting with just 10\% labeled data, our model maintains performance comparable to some of the general contrastive SSL methods. These results strongly support the data efficiency and robustness of our approach. By integrating time- and frequency-domain information, our model delivers reliable and accurate predictions with minimal supervision, making it highly applicable to real-world neuroimaging scenarios.

    \begin{table*}[t!]
    \centering
    \scriptsize  % 缩小字体
    \caption{Ablation study performance (\%) for frequency information on ABIDE and ADHD datasets. The best results are bold, and the second results are underlined.}
    \label{tab: ablation}
    \setlength{\extrarowheight}{2pt} 
    \resizebox{\linewidth}{!}{  % 缩放表格以适应页面宽度
    \begin{tabular}{c|cccc|cccc}
        \hline
        \multirow{2}{*}{\textbf{Methods}}  &  \multicolumn{4}{c|}{\textbf{ABIDE}} & \multicolumn{4}{c}{\textbf{ADHD}} \\
        \cline{2-9}
        & \textbf{ACC} & \textbf{AUC}  & \textbf{Recall}  & \textbf{F1-score} &  \textbf{ACC} & \textbf{AUC}  & \textbf{Recall}  & \textbf{F1-score} \\
        \hline
        \textcolor{black}{FENet\_T}   & $ \textcolor{black}{55.0\pm 2.7}$ & $ \textcolor{black}{56.1 \pm 1.3}$ & $ \textcolor{black}{59.9 \pm 2.8}$ & $ \textcolor{black}{57.3 \pm 2.8}$ & $ \textcolor{black}{57.3 \pm 4.3}$ & $ \textcolor{black}{60.5 \pm 4.5}$ & $ \textcolor{black}{61.8 \pm 3.4}$ & $ \textcolor{black}{60.3 \pm 4.2}$ \\
        \textcolor{black}{FENet\_F} & $ \textcolor{black}{56.7\pm 3.3}$ & $ \textcolor{black}{57.0 \pm 2.7}$ & $ \textcolor{black}{57.8 \pm 7.9}$ & $ \textcolor{black}{\mathbf{64.5 \pm 2.9}}$ & $ \textcolor{black}{{59.8 \pm 4.2}}$ & $ \textcolor{black}{{61.2 \pm 4.6}}$ & $ \textcolor{black}{{62.4 \pm 4.7}}$ & $ \textcolor{black}{61.9 \pm 5.7}$ \\
        \hline
        FENet\_FL &$\underline{58.4\pm 6.2}$ & $\underline{58.2 \pm 4.5}$ & $\underline{58.6 \pm 4.6}$ & $59.3 \pm 5.1$  &$\underline{60.6\pm 3.8}$ & $\underline{62.7 \pm 3.1}$ & ${62.1 \pm 3.5}$ & $\underline{62.9 \pm 6.2}$ \\
        FENet\_FM &$56.4\pm 4.5$ & $57.4 \pm 4.3$ & $58.1 \pm 5.1$ & $58.4 \pm 5.3$  &$58.3\pm 5.2$ & $61.4 \pm 4.2$ & $\underline{62.8 \pm 4.7}$ & ${60.4 \pm 4.6}$ \\
        FENet\_FH & $\mathbf{59.3\pm 5.6}$ & $\mathbf{60.3\pm 4.6}$ & $\mathbf{59.8\pm 5.7}$ & $\underline{62.5\pm 7.3}$ &$\mathbf{63.3 \pm 4.3}$ &$\mathbf{64.2 \pm 4.2}$ &$\mathbf{63.5 \pm 7.4}$ &$\mathbf{63.5 \pm 6.5}$\\
        \hline
    \end{tabular}
    }
    \end{table*}

    \subsection{Ablation Study}
    \subsubsection{Effectiveness of Frequency Information}

	\textcolor{black}{Our main contribution lies in systematically incorporating frequency information into the representation learning framework to investigate its role in brain disorder detection, with a particular emphasis on the diagnostic value of high-frequency components. To this end, we conducted a comprehensive evaluation of the effectiveness of joint time–frequency representations and analyzed model performance under different frequency conditions. To disentangle the contributions of time-domain and frequency-domain information, we designed two controlled model variants. Specifically, we first applied random feature perturbations to the brain graphs to ensure robustness. For the FENet\_T variant, only the TGNN is used during the SSL phase, and the resulting time-domain representations are employed for downstream classification. Conversely, the FENet\_F variant uses only the FGNN to generate frequency-based representations for classification. As shown in the first two rows of Table~\ref{tab: ablation}, both FENet\_T and FENet\_F outperform most of the general contrastive SSL baselines, with FENet\_F achieving better performance. This demonstrates that frequency features indeed encode disease-relevant information, while also improving the time-domain representations under the guidance of the time–frequency consistency concept.} To further explore the impact of different frequency bands, we introduced three filtered variants: FENet\_FH (high-pass), FENet\_FM (band-pass), and FENet\_FL (low-pass). As shown in the last three rows of Table~\ref{tab: ablation}, all three variants outperform the time-only FENet\_T baseline, indicating the benefit of incorporating frequency information. Among them, FENet\_FH achieves the best performance on both ABIDE and ADHD datasets, suggesting that high-frequency components often capture rapid neural oscillations induced by task exposure, which are closely linked to cognitive processing deficits associated with psychiatric disorders~\cite{sasai2021frequency}. In contrast, the mid-frequency variant FENet\_FM contributes less to classification performance and even underperforms the full-spectrum FENet\_F model, implying that these bands may carry limited disorder-specific information. On the other hand, FENet\_FL also yields significantly improved results, supporting the understanding that low-frequency signals reflect global brain dynamics and long-range functional connectivity, which are highly informative for detecting psychiatric disorders.

    These findings are consistent with established medical literature \cite{huang2016graph, sasai2021frequency}, reinforcing the importance of jointly considering both low-frequency and high-frequency information in fMRI-based diagnostic modeling. Motivated by these insights, our approach integrates both low-pass and high-pass filters to extract complementary low- and high-frequency components while discarding potentially noisy or redundant mid-frequency information. This dual-filter strategy enables efficient and focused frequency-domain analysis, ultimately contributing to the superior detection performance observed on both datasets.

       \begin{figure}[t]
        \centering
        \begin{subfigure}[b]{0.45\textwidth} % 调整宽度
            \centering
            \includegraphics[width=\textwidth]{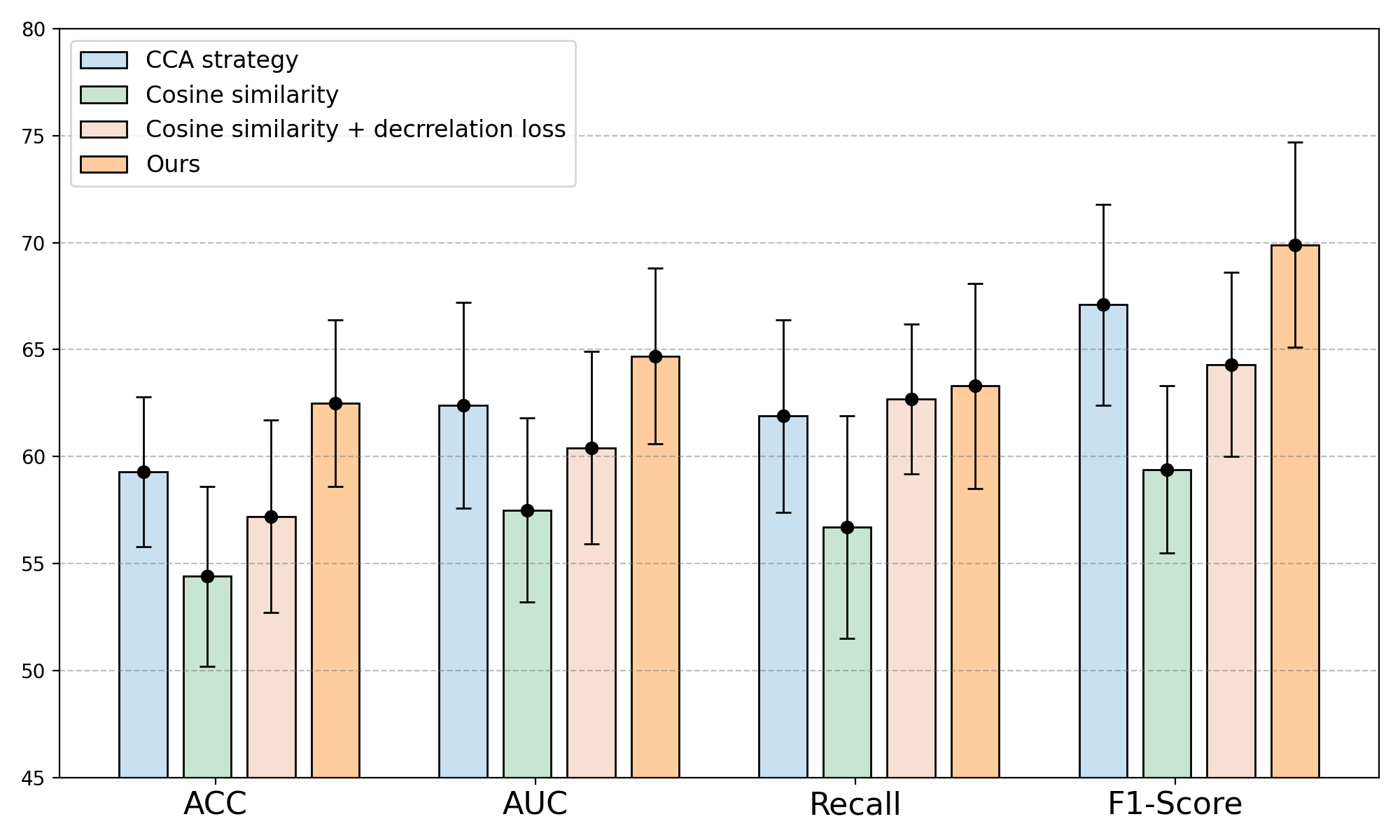}
            \caption{ABIDE}
        \end{subfigure}
        \hspace{5pt}
        \begin{subfigure}[b]{0.45\textwidth} % 调整宽度
            \centering
            \includegraphics[width=\textwidth]{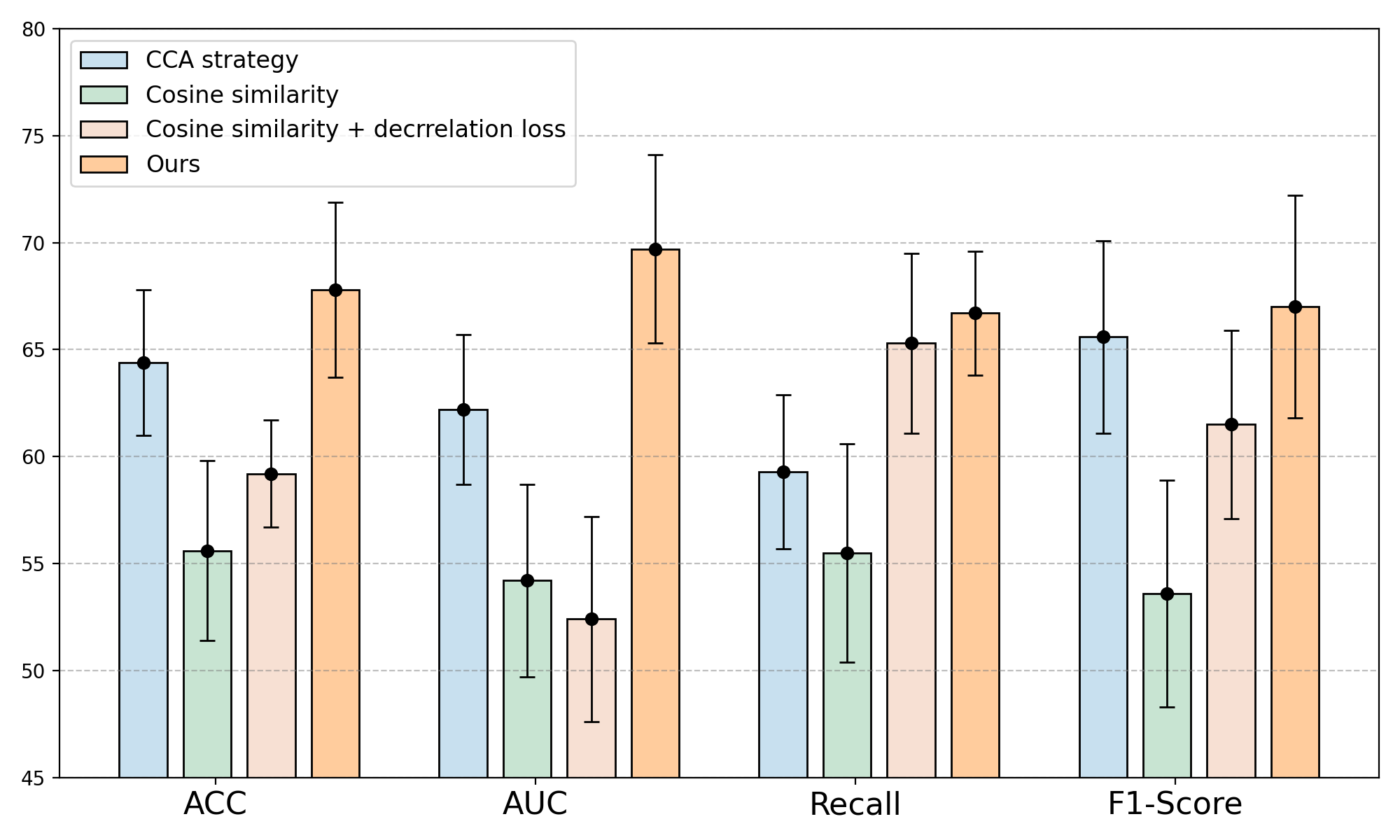}
            \caption{ADHD}
        \end{subfigure}

        \caption{Ablation study performance (\%) for different learning objective settings. The vertical lines (error bars) on the graph represent the variance of the data.}
        \label{fig:ablation_result2}
        \end{figure}

    \subsubsection{Effectiveness of Learning Objective}
    \textcolor{black}{Our learning objective is fundamentally designed to encourage cross-domain representation alignment under the theoretical guidance of domain consistency, while simultaneously reducing information redundancy through decorrelation. In line with standard similarity-based SSL baselines~\cite{thakoor2021bootstrapped}, we adopt cosine similarity as an alternative to our CCA-based strategy for measuring representation similarity between the two views. Additionally, we design an additional variant to analyze the importance of the decorrelation term.} Specifically, we consider the following three configurations:
    (i) CCA strategy: a variant using CCA but with identical trade-off coefficients applied to representations from both domains;
    (ii) Cosine Similarity: replacing our CCA-based objective with a cosine similarity strategy following;
    (iii) Cosine Similarity + decorrelation loss: an improved version of the cosine similarity approach, enhanced by incorporating an additional decorrelation term to examine its contribution. 
    
    The experimental results, presented in Figure~\ref{fig:ablation_result2}, reveal several key insights. First, the CCA strategy consistently outperforms the pure cosine similarity baseline across all evaluation metrics, underscoring the benefit of explicit alignment constraints for cross-domain representation fusion. However, this variant still falls short of our final model. We attribute this performance gap to the assumption of equal importance across domains, an oversimplification that disregards the heterogeneous discriminative capacity of time- and frequency-domain features. Our method addresses this by applying domain-specific trade-off coefficients, allowing the model to better balance and leverage complementary signals from both domains. 
    Moreover, incorporating a decorrelation term significantly improves the performance of the cosine similarity-based variant. This confirms the importance of preserving domain-specific semantic structures and avoiding over-alignment, which can otherwise obscure meaningful features unique to each view. Nevertheless, despite these improvements, cosine similarity-based methods still lag behind CCA-based approaches. This is likely due to the lack of global alignment constraints in cosine similarity, making the model more susceptible to training instability, representational collapse, or convergence to suboptimal local minima. 
    In summary, these ablation results validate the effectiveness of our proposed learning objective. They demonstrate that our domain-consistency-guided CCA formulation strikes a superior balance between alignment and distinctiveness, enabling the model to learn more robust, semantically rich, well-aligned time-frequency representations.

        \begin{figure}[t]
        \centering
        \begin{subfigure}[b]{0.45\textwidth} % 调整宽度
            \centering
            \includegraphics[width=\textwidth]{ 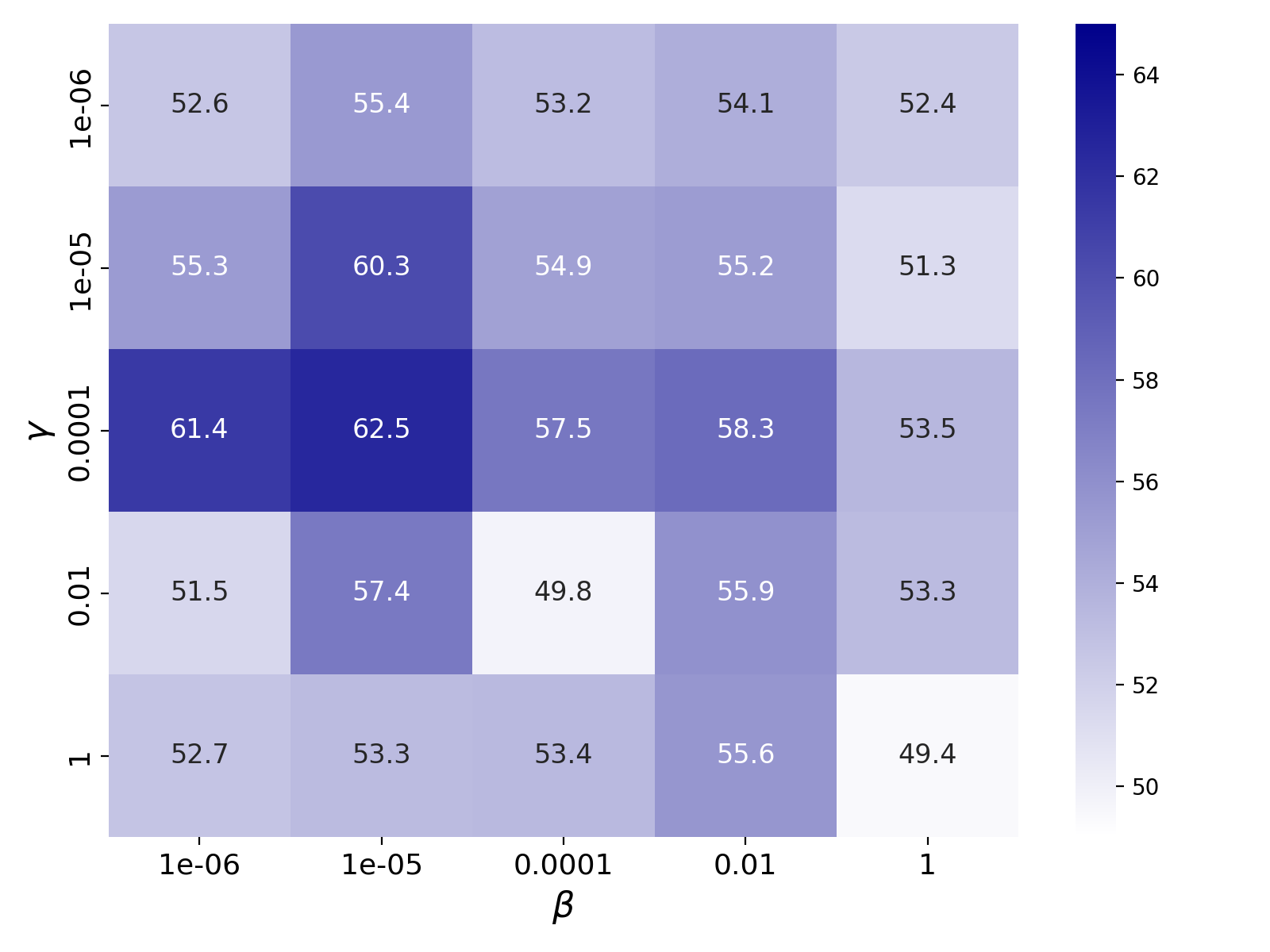}
            \caption{ABIDE}
        \end{subfigure}
        \hspace{5pt}
        \begin{subfigure}[b]{0.45\textwidth} % 调整宽度
            \centering
            \includegraphics[width=\textwidth]{ 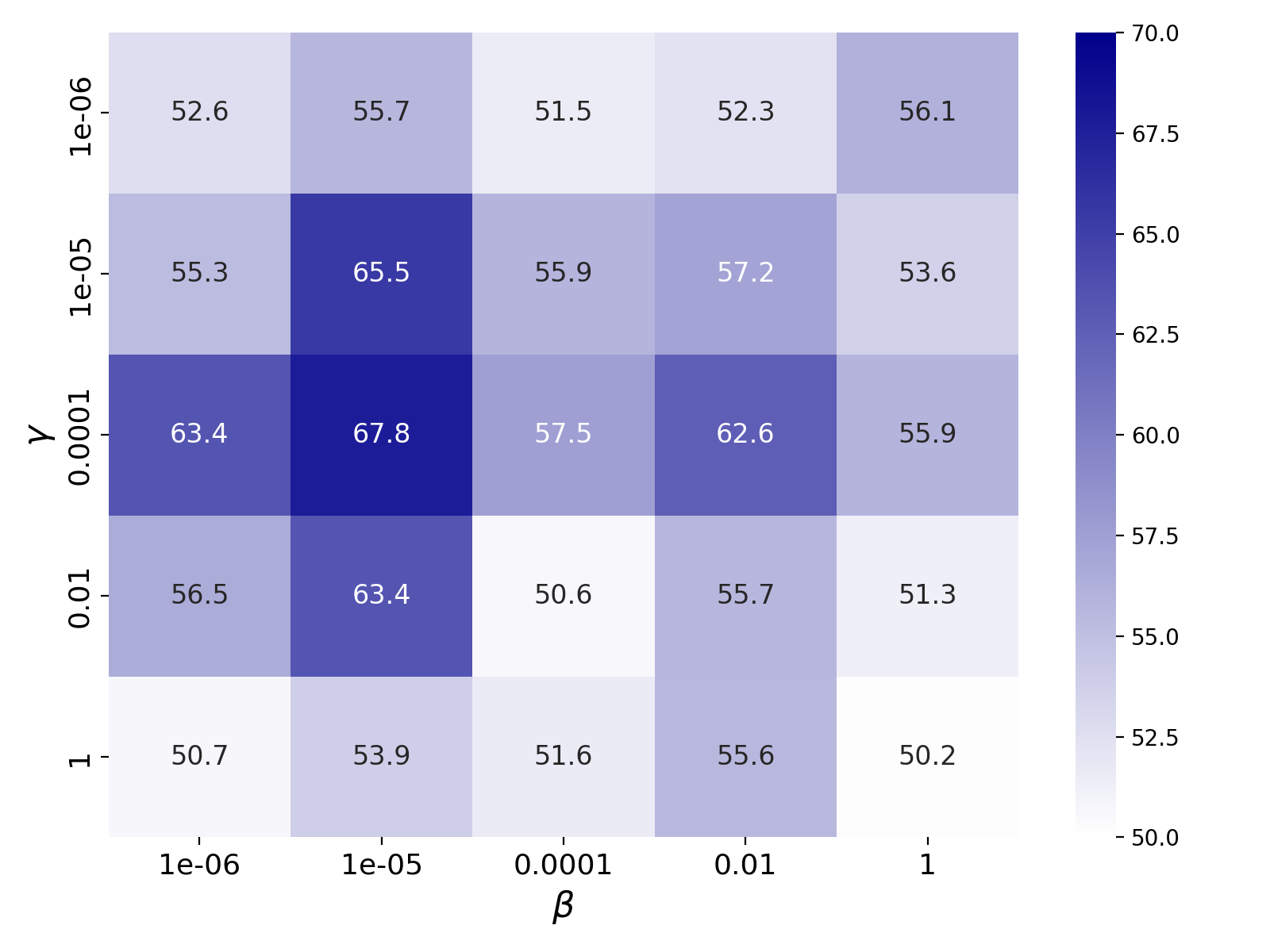}
            \caption{ADHD}
        \end{subfigure}

        \caption{\textcolor{black}{Accuracy (\%) of FENet with different values of $\gamma$ and $\beta$ in Equation \eqref{Loss}.}}
        \label{fig:ablation_result3}
        \end{figure}
        
    \subsubsection{Effectiveness of the Decorrelation Intensity Term}

    To achieve an optimal trade-off between time-domain and frequency-domain information, we conduct a comprehensive sensitivity analysis by varying the decorrelation parameters $\gamma$ and $\beta$ in the objective function, which respectively control the influence of time-domain and frequency-domain regularization terms. This analysis is crucial to evaluate the individual and joint contributions of different information sources in fMRI data to the performance of the model. As illustrated in Figure~\ref{fig:ablation_result3}, the results across both datasets show that the model performs consistently well within a narrow parameter range ($\gamma \leq 1e^{-5}, \beta \leq 1e^{-4}$), beyond which performance begins to degrade. We attribute this stability to the model ability to adhere to the principle of domain consistency, which encourages the learning of fusion representations that capture shared and invariant features across time and frequency domains. 
    In contrast, when either parameter is set too high, the model becomes overly biased toward decorrelating representations, which may suppress task-relevant semantic information. As a result, the learned embeddings drift from discriminative space to less informative subspaces, weakening the model representational capacity. 
    Interestingly, we observe that the optimal value for the frequency-domain parameter $\beta$ is relatively more significant than that of the time-domain parameter $\gamma$, suggesting that frequency-domain representations inherently exhibit greater redundancy and benefit more from decorrelation. This further implies that frequency-domain signals contain richer and more invariant dynamic patterns, which are complementary and vital in enhancing the integrated brain network representation. 
    Based on this analysis, we empirically set $\gamma = 1e^{-5}$ and $\beta = 1e^{-4}$ for all subsequent experiments to ensure a balanced and effective utilization of information from both domains, promoting robust and semantically meaningful brain graph representations.

\section{Conclusion}
\label{Conclusion}
This paper introduces FENet, an innovative and effective SSL framework designed to learn diverse information about fMRI data under limited-sample conditions, facilitating efficient disease detection. FENet explicitly integrates frequency and time information to provide a holistic understanding of brain activities. Specifically, multi-view brain graphs in both time and frequency domains are constructed, and corresponding information encoders in two domains are employed to extract temporal-spectral features while exploring their correlations with brain function. Further, to balance the utilization of these features, FENet uses a %TF-C-based 
domain consistency-guided learning objective to maximize the invariance across two views, constructing frequency-enhanced brain network representations. Experimental results on two real-world neural imaging datasets demonstrate that FENet significantly outperforms existing baselines. Moreover, our analysis highlights that frequency information provides critical insights into disease-related brain dynamics, with high-frequency components playing a particularly significant role in capturing localized disruptions and neural oscillation abnormalities associated with psychiatric disorders.

In future research, we will further enhance the modeling of temporal dependencies across multiple domains. Additionally, we aim to design an adaptive representation fusion module that leverages frequency-domain information, enabling a more effective joint analysis of time-domain and frequency-domain features. \textcolor{black}{While our method was evaluated separately on ABIDE and ADHD-200 datasets, cross-database analysis remains challenging due to differences in acquisition protocols. Future work could explore harmonization techniques and domain adaptation approaches to enable robust cross-dataset generalization, as well as to facilitate the discovery of differences and shared biomarkers across related psychiatric disorders.}

\bibliographystyle{ACM-Reference-Format}
\bibliography{FENet}

\end{document}